\documentclass[12pt]{article}

\newcommand{\bbr}{I\!\! R}

\newcommand{\2}{$^2$}
\newcommand{\3}{$^3$}
\newcommand{\4}{$_4$}
\newcommand{\5}{$_5$}
\newcommand{\x}{arXiv:}

\usepackage{graphicx}

\begin{document}
\thispagestyle{empty}
\begin{center}

\null \vskip-1truecm \vskip2truecm {\bf Quintessential
Maldacena-Maoz Cosmologies\\} \vskip1truecm Brett McInnes
\vskip1truecm

 National University of Singapore

email: matmcinn@nus.edu.sg\\

\end{center}
\vskip1truecm \centerline{ABSTRACT} \baselineskip=15pt
\medskip

Maldacena and Maoz have proposed a new approach to holographic
cosmology based on Euclidean manifolds with
 disconnected boundaries. This approach appears, however, to be in conflict with the known geometric
 results [the Witten-Yau theorem and its extensions] on spaces with boundaries of non-negative scalar
  curvature. We show precisely how the Maldacena-Maoz approach evades these theorems. We also
  exhibit Maldacena-Maoz cosmologies with [cosmologically] more
  natural matter content, namely quintessence instead of
  Yang-Mills fields, thereby demonstrating that these cosmologies
  do not depend on a special choice of matter to split the
  Euclidean boundary. We conclude that if our Universe is fundamentally \emph{anti-de Sitter-like}
  [with the current acceleration being only temporary], then this may force us to confront the holography
   of spaces with a connected
   bulk but a disconnected boundary.
\vskip3.5truecm
\begin{center}

\end{center}

\newpage

\addtocounter{section}{1}
\section*{1. The Holography of a Crunch }
A theoretical understanding of the expansion history of the
Universe should illuminate two fundamental aspects. The first, of
course, is the acceleration [see for example
\cite{kn:carroll1}\cite{kn:linder}], interpreted theoretically
\cite{kn:larsen} in terms of ``de Sitter-like" physics. The second
is the possibility that the de Sitter state is metastable, and
will ultimately be succeeded by an ``anti-de Sitter-like" state
terminating in a Big Crunch. The thorny problems associated with
the holographic \cite{kn:bousso} picture of de Sitter spacetime
would then be replaced by a new set of challenges: what is the
holographic description of an anti-de Sitter Crunch?

There are in fact observational hints
\cite{kn:sahni}\cite{kn:nesseris} raising questions as to whether
the Universe has simply evolved from a matter-dominated condition
directly to the current vacuum-dominated state; there is some
evidence that the evolution has been considerably more interesting
than that. A future transition from acceleration to collapse is
therefore not as implausible as it may seem from an observational
point of view.

From a theoretical point of view, it has long been known
\cite{kn:dine} that there are arguments which lead to the
conclusion that if a de Sitter phase can be realised in string
theory at all, this phase can only be metastable. This has been
emphasised again in recent work on the cosmological constant
problem, for example in \cite{kn:linde}\cite{kn:becker}. The point
has been explained very simply in \cite{kn:giddings}, where it is
emphasised that, in a theory with extra dimensions controlled by a
radial dilaton, the potential must \emph{vanish at infinity}
except for very exotic matter fields. The vanishing of the
potential at infinity implies that a de Sitter equilibrium cannot
correspond to a global
 minimum, and this leads either to an eventual catastrophic decompactification [if the potential remains positive]
 or, perhaps more plausibly, to a transition to contraction
 culminating in a Big Crunch. [The
  only exception to the statement that the potential vanishes at infinity would
  be given by
   ``phantom" matter \cite{kn:caldwell} with an
  equation-of-state parameter below $-$2. In view of the recent data
  analyses supporting phantom cosmologies --- see for example
  \cite{kn:sahni}\cite{kn:pad} --- this should be investigated: note
  that while the observational data exclude such low values for the
  \emph{total} equation-of-state parameter, they do not rule out a
  mixture of such exotic matter with more normal varieties. But it is
  certainly not known how to obtain such matter in the string context,
  so we shall not consider this possibility further.]

The general thesis underlying this ``Crunchy" view of cosmic
evolution is that our Universe is \emph{fundamentally anti-de
Sitter-like rather than de Sitter-like}: the current acceleration
is just a passing phase which does not dictate our ultimate fate
\cite{kn:linde2}. That is, the structure of infinity is to be
understood in terms of asymptotically anti-de Sitter rather than
asymptotically de Sitter spacetimes. Since anti-de Sitter-like
cosmologies generically have Bangs and Crunches, the consequences
for ``holographic cosmology" are obviously profound.

``Anti-de Sitter-like" cosmologies are, by definition, obtained by
introducing matter into AdS\4 and allowing it to distort the
geometry. The study of such cosmologies from a holographic point
of view was recently initiated by Maldacena and Maoz
\cite{kn:maldacena}, who point out that anti-de Sitter-like
cosmologies correspond to Euclidean manifolds with a conformal
compactification such that the boundary consists of two
disconnected components. This immediately opens the way to the use
of suitably generalised AdS/CFT techniques, and one might well
hope to investigate holographic anti-de Sitter cosmology, and
possibly the transition from acceleration to collapse, in this
way. [The use of spaces with multiple boundary components to
generate \emph{de Sitter}-like cosmologies was explored in
\cite{kn:bigsmash}. Thus these spaces may be relevant to
\emph{both} of the rival candidates for holographic theories of
cosmology.]

However, it is well known that Witten and Yau \cite{kn:yau} have
shown that such Euclidean spacetimes give rise to badly behaved
field theories on the boundary if the bulk is a geodesically
complete Einstein manifold of negative scalar curvature. Maldacena
and Maoz avoid this problem by allowing the bulk matter to act on
the bulk geometry, so that the bulk metric is no longer Einstein.
In essence, the key to understanding cosmological evolution from
the AdS/CFT point of view is to take into account this
back-reaction, moving beyond treating the bulk as a fixed
background.

The prospect of using holography in cosmology is enticing, but it
raises many questions. Witten and Yau actually claim that their
result still holds even for some bulk manifolds which are
\emph{not} Einstein manifolds. It follows that the Maldacena-Maoz
cosmologies require bulk matter of some specific kind --- it is
not enough merely to introduce arbitrary forms of matter, which
might still be governed by this more general version of the
Witten-Yau theorem. We must therefore ask: what specific
properties of the configurations considered by Maldacena and Maoz
allow them to avoid the instabilities discussed by Witten and Yau?
Furthermore, in their effort to obtain well-behaved field theories
on the boundary, Maldacena and Maoz are led to use bulk matter of
a kind [Yang-Mills fields] which is not normally considered to be
suitable for cosmology. In particular, this matter satisfies the
Strong Energy Condition at all times and cannot describe either
the current acceleration or of course the subsequent transition to
collapse.

These points might lead one to suspect that the splitting of the
Euclidean boundary could still be avoided if cosmologically more
familiar matter were used instead of the special Yang-Mills
configurations considered in \cite{kn:maldacena}. Our objective
here is to show that this is \emph{not} the case. We introduce a
one-parameter family of cosmological models obtained by inserting
\emph{quintessence} [see for example \cite{kn:ratra}] into AdS\4,
instead of Yang-Mills fields. These are Bang/Crunch cosmologies
which nevertheless have temporarily accelerating phases; they
therefore yield a very simple model of the transition from
acceleration to collapse. Furthermore, the Euclidean version has a
disconnected boundary, precisely as in \cite{kn:maldacena}. Using
these, we can explain precisely what properties bulk matter should
have in order to evade the Witten-Yau theorem. [We are not
claiming to have solved the difficult problem of obtaining
quintessence from string or M-theory; we must assume that this is
possible, perhaps along the lines indicated in \cite{kn:townsend}
or \cite{kn:ish1}\cite{kn:jarv}. If that can indeed be done, then
one expects the quintessence to have a well-behaved description in
terms of a field theory configuration on the Euclidean boundary.]

Our conclusion is rather surprising: it is actually quite easy to
avoid the strictures of the Witten-Yau theorem, even in its
strongest version [due to Cai and Galloway \cite{kn:galloway}]. We
conclude that cosmological models with Euclidean versions having
multiple boundaries [henceforth, ``Maldacena-Maoz cosmologies"] do
not in general lead to unacceptable physics. Furthermore, they are
``generic" in the sense that they do \emph{not} require the use of
the interesting but [in the cosmological
  context] somewhat unusual bulk matter studied in
\cite{kn:maldacena}. They therefore force us to confront the
apparent conflict with holography which arises when one apparently
has \emph{two} independent field theories associated with
\emph{one} bulk. This truly fundamental puzzle cannot, in short,
be disposed of by claiming that it cannot arise in physically
realistic circumstances.
\bigskip

[We note before proceeding that the reader should not confuse the
wormholes considered by Maldacena and Maoz with Lorentzian
wormholes \cite{kn:visser}. Only
 the Euclidean versions of the Maldacena-Maoz spaces have
 wormholes. They also differ from the AdS wormholes studied in \cite{kn:adsworm}, which have the local
 geometry of [Euclidean] AdS itself, and which have to be sustained by a brane
 at the wormhole throat; though \cite{kn:adsworm} was also
 motivated by a wish to investigate the ``disconnected boundary"
 problem.]

\addtocounter{section}{1}
\section*{2. Anti-de Sitter Spacetime and its Crunchy Relatives}

The Maldacena-Maoz cosmologies, and their rather subtle
relationship with anti-de Sitter spacetime itself, can be
understood with the help of the points raised in the following
discussion.

Anti-de Sitter spacetime is of course not very interesting as a
cosmological model. Written in FRW form, its metric is [in four
dimensions] given by
\begin{equation}\label{eq:A}
g^-(AdS_4) =  - dt^2 + \textup{cos}^2(t/\textup{L})[dr^2 +
\textup{L}^2 \textup{sinh}^2(r/\textup{L})[d\theta^2 +
\textup{sin}^2(\theta)d\phi^2]],
\end{equation}
where the cosmological constant is $-$3/L\2, and where a negative
superscript will always indicate a Lorentzian metric, a positive
superscript denoting a Euclidean metric. Notice that the spatial
sections are just copies of the three-dimensional hyperbolic space
H\3 of sectional curvature $-$1/L\2, with the metric expressed in
terms of polar coordinates. Thus (\ref{eq:A}) \emph{appears} to
represent a Bang/Crunch cosmology with negatively curved spatial
sections: a strange mixture of the traditional ``closed" and
``open" cosmological models. It also \emph{appears} to represent a
time-dependent geometry, as is normal in cosmology. In reality,
the apparent spacelike singularities at t = $\pm\pi$/2 are mere
coordinate singularities, which arise because all of the timelike
geodesics perpendicular to the spatial surface at t = 0 intersect
periodically. The apparent time-dependence is likewise illusory,
since the full anti-de Sitter geometry has a timelike Killing
vector, which may be thought of as arising from the Killing
spinors associated with the AdS supersymmetries. [This is the
counterpart of the fact that the de Sitter spacetime, which has no
timelike Killing vector, can be made to appear static by means of
a choice of coordinates.] See \cite{kn:gibbons} for a good
discussion of these peculiarities of anti-de Sitter spacetime.

The reason that AdS\4 can, despite appearances, avoid being a
Bang/Crunch spacetime, is essentially that it contains nothing
apart from the matter which supplies the [negative] cosmological
constant --- this would be p-form matter in the string theory
context. This simplifies the structure of the curvature tensor to
the extent that AdS\4 fails to satisfy the \emph{generic
condition} in the Hawking-Penrose cosmological singularity
 theorem [\cite{kn:hawking}, page 266]. We can therefore expect that the introduction of matter into AdS\4 will
 cause the geometry to satisfy the generic condition and so produce a singular cosmological spacetime of the
  kind we are seeking, since AdS\4 does satisfy all of the other conditions of the singularity theorem,
  \emph{including} the Strong Energy Condition. [Strictly speaking, it does not have
  a compact edgeless achronal set as the relevant version of the
  singularity theorem requires, but by taking the quotient by a
  freely acting group which compactifies the spatial sections, we
  obtain a spacetime which does have such sets; this spacetime is
  still non-singular, so we see that it is indeed the generic
  condition which is the relevant one here. Notice that anti-de Sitter spacetime
  differs in this regard from de Sitter spacetime, which
  violates \emph{two} conditions
   of the Hawking-Penrose theorem, namely the generic condition \emph{and} the Strong Energy Condition.
    Thus the introduction of small amounts of matter into \emph{de Sitter} spacetime should
     not be expected to cause the spacetime to become singular.] We now consider a simple example illustrating
      this point. [For a very different approach to obtaining
      cosmological spacetimes from anti-de Sitter spacetime, see
      \cite{kn:cvetic}.]

The current observational evidence  does not rule out negatively
curved spatial sections, but the sections are close to being flat.
Let us take the AdS\4 metric and simply replace the negatively
curved spatial sections with flat ones. For reasons which will
become apparent, we take the flat sections to be compact, [say]
cubic tori. For Bang/Crunch cosmologies there will be a toral
spatial section of maximum size [at t = 0]; we shall specify that
the circumferences of that torus shall be 2$\pi$A, for some
suitably large constant A. [Other compact flat manifolds are
equally acceptable, though of course one may prefer to impose
orientability.]

Modifying the AdS\4 metric in this way, we obtain
\begin{equation}\label{eq:B}
g^-(1,A) = -dt^2 + A^2 \textup{cos}^2(t/\textup{L})[d\theta_1^2 +
d\theta_2^2 + d\theta_3^2],
\end{equation}
where the torus is parametrised by angles and where the notation
$g^-(1,A)$ will be explained below. Unlike the anti-de Sitter
metric, this metric is genuinely singular: it has a Big Bang at t
= $-\pi$/2 and a Big Crunch at t = $+\pi$/2, as we shall prove
explicitly later. Of course, it is not like AdS\4, which solves
the Einstein equation with no matter apart from that which
generates the negative cosmological constant: we have introduced
matter [of a kind to be described below] into an anti-de Sitter
background. The effect of this matter is to flatten the spatial
sections, to introduce spacelike singularities at t = $\pm\pi$/2,
and also to remove the timelike Killing vector. [This last follows
from the fact, to be established below, that the coordinates in
(\ref{eq:B}) cover the entire spacetime.] One can understand this
physically by thinking of the \emph{negative} cosmological
constant as being associated with an ``attractive force" which
increases with separation. As soon as we introduce matter into
AdS\4, this ``attraction" inevitably results in a Crunch. In this
sense, $g^-(1,A)$ is ``more generic" than the pure anti-de Sitter
metric. [Like AdS\4, this spacetime satisfies the Strong Energy
Condition [see below], but it also satisfies the generic condition
and it has compact achronal edgeless sets because the spatial
sections are compact; and so it has to be singular by the
Hawking-Penrose theorem.]

Now the Euclidean version of AdS\4 is of course the hyperbolic
space H$^4$, the four-dimensional simply connected space
 of constant negative curvature. As is well known from studies of the AdS/CFT correspondence \cite{kn:witten}, the
 conformal boundary of H$^4$ is a conformal three-sphere, S\3, which is compact and connected. When however we
  consider the Euclidean version of $g^-(1,A)$, given by
\begin{equation}\label{eq:C}
g^+(1, A) = dt^2 + A^2 \textup{cosh}^2(t/\textup{L})[d\theta_1^2 +
d\theta_2^2 + d\theta_3^2],
\end{equation}
we see immediately that, at least in the most obvious
interpretation, the conformal boundary of the underlying manifold
is compact but \emph{not} connected: it consists of two copies of
the torus, T\3. The introduction of matter into AdS\4 has not just
flattened the conformal boundary: in the Euclidean picture
\emph{it has split it into two connected components.} [We chose
the flat sections to be compact so that the conformal boundary
should be compact, thereby avoiding all of the complications which
arise in AdS/CFT if the boundary is allowed to be non-compact.
This also has the benefit of making it clear that the boundary is
indeed disconnected --- this is sometimes far from evident when
the sections are non-compact, as for example in the foliation of
AdS\5 by AdS\4 slices. [See the discussion in the Conclusion,
however.] We stress that any compact flat three-dimensional
manifold would be as suitable as the cubic torus we are using
here.]

Maldacena and Maoz \cite{kn:maldacena} propose a more general
version of this construction as a way of understanding Bang/Crunch
cosmologies from the AdS/CFT point of view. They propose to study
the holography
 of general Euclidean manifolds of the form
\begin{equation}\label{eq:D}
g^+(\textup{F}, \Sigma) = dt^2 + \textup{F}^2(t)\;g(\Sigma),
\end{equation}
where F is a nowhere-zero function which resembles
e$^{|t|/\textup{L}}$ as t tends to $\pm\infty$ [see Section 6 for
a precise version of this], where L is some positive constant, and
 where $\Sigma$ is some compact three-manifold with Euclidean metric $g(\Sigma)$. Such manifolds have
 Bang/Crunch cosmologies as their Lorentzian versions, while the Euclidean version locally resembles Euclidean
  anti-de Sitter spacetime near t = $\pm\infty$. [That is, the sectional curvatures all asymptotically
   approach $-$1/L\2; this is true whatever the geometry of $\Sigma$ may be.] Clearly there is an
   opportunity to bring AdS/CFT techniques to bear on Bang/Crunch cosmologies in this way.
   However, it is also clear that the conformal compactification of the Euclidean version has a
   boundary which consists of \emph{two} copies of the compact manifold $\Sigma$.

This suggestion, therefore, immediately forces us to confront one
of
 the deepest problems in holography: how does the holographic
 philosophy deal with a situation in which there are \emph{two}
 boundaries, inhabited by two [presumably] distinct field theories,
 but only one bulk? The correlators between the two boundaries should
 factor from the point of view of the field theory,
 but not from the point of view of the bulk: a flagrant violation of the holographic
 principle. This serious problem was pointed out by
 Witten and Yau \cite{kn:yau}, who suggested an ingenious solution which we shall now explain.

\addtocounter{section}{1}
\section*{3. The Witten-Yau Theorem}

Witten and Yau proposed to deal with the problem of disconnected
boundaries in the most radical way, by attempting to prove that
such a situation cannot arise in a
 physically reasonable manner. They showed that

\medskip

[a] if the bulk is a geodesically complete connected Euclidean
Einstein manifold of
 negative scalar curvature, and

\medskip
[b] if the conformal structure induced on any component of the
boundary is represented by a metric of
 positive constant scalar curvature,

\medskip
then the conformal boundary \emph{must be connected}. [Simplified
proofs, with many related results, were given in
\cite{kn:galloway} and \cite{kn:anderson} [see also
\cite{kn:anderson2}]; another proof, with the slightly stronger
hypothesis that the scalar curvature should
 be positive on \emph{all} components of the boundary,
 was given in \cite{kn:wang}.]

The condition that the scalar curvature on the boundary should be
positive can be partly justified by noting that, in the negative
case, there is a non-perturbative instability arising from the
nucleation of branes in the bulk; negative scalar curvature at
infinity implies that the action is decreased as the brane moves
towards the boundary \cite{kn:seiberg}. [For a recent discussion
of this, and of the difficulties which arise when one attempts to
suppress the instability, see \cite{kn:buchel}.] As it stands the
theorem does not explain what happens in the case where the
boundary has zero scalar curvature; indeed, in general the
physical acceptability of this case is not fully understood, but
certainly we must allow the special case in which the boundary is
completely flat [and not just scalar-flat]. Fortunately the
Witten-Yau theorem was improved by Cai and Galloway
\cite{kn:galloway} so that the same conclusion can be reached but
with the condition on the boundary scalar curvature weakened to
``non-negative" instead of ``positive". The upshot is that, in
physically reasonable cases, the boundary \emph{cannot} be
disconnected if the bulk is a complete Euclidean Einstein manifold
of negative scalar curvature. Thus the ``two boundaries/one bulk"
conundrum cannot arise in that case.

Taking the bulk to be an Einstein manifold means that we are
ignoring the effect of bulk matter on the bulk geometry. That is a
good approximation in some circumstances, but not in all ---
certainly not in cosmology. So for anti-de Sitter-like cosmologies
the question returns: can there be non-perturbatively stable
spacetimes with two boundaries and one bulk if the bulk matter is
allowed to act on the bulk metric, so that it is no longer an
Einstein metric? Witten and Yau suggested an answer in this case
also. They argued that their result continues to hold if the
Einstein condition is weakened in the following way. Think of the
Ricci tensor as a (1,1) tensor, so that its eigenvalues are
well-defined; they are functions of position in general. If the
eigenvalue functions are Ric$_{(j)}$, where j ranges from 0 to 3
in four dimensions, then the Einstein condition is just
\begin{equation}\label{eq:E}
\textup{Ric}_{(j)} = -3/\textup{L}^2, \;\;\; j\;=\;0,1,2,3.
\end{equation}
Witten and Yau weaken this to the condition that the eigenvalues
of the Ricci tensor should be bounded below \emph{everywhere in
the bulk}  by their asymptotic values as the boundary is
approached. If the asymptotic sectional curvature is $-$1/L\2,
then the condition replacing (\ref{eq:E}) is just
\begin{equation}\label{eq:F}
\textup{Ric}_{(j)} \geq -3/\textup{L}^2 \;\;\;j\;=\;0,1,2,3.
\end{equation}
In short, the boundary still has to be connected as long as the
back-reaction of bulk matter always tends to \emph{increase} the
Ricci curvature. [Allowing the Ricci eigenvalues to become
functions of position, however, immediately raises questions as to
what exactly happens to these functions as infinity is approached.
This subtle point was raised by Cai and
  Galloway \cite{kn:galloway}, who stressed the importance of the
  \emph{rate} at which the Ricci eigenvalues approach the asymptotic
  values. We shall explain this in detail
  below.]

Witten and Yau state that their condition on the Ricci curvature
corresponds to having matter fields excited in an asymptotically
anti-de Sitter space. [The asymptotic values of the Ricci
eigenvalues
  are interpreted as the anti-de Sitter cosmological constant.]  The
first question is then: what kind of matter would correspond to a
geometry with such Ricci eigenvalues? Secondly we should ask: is
it physically reasonable to impose the conditions demanded by Cai
and Galloway
 on the asymptotic data? For an asymptotically anti-de Sitter black
 hole, there are well-motivated conditions on the rate at which
the metric should approach the AdS\4 metric \cite{kn:magnon}, but
in general one would not expect black hole boundary conditions to
be relevant to cosmology.

In the examples considered by Maldacena and Maoz, the bulk matter
is typically a Yang-Mills instanton or meron, and, precisely in
order to evade the ``Einstein" version of the Witten-Yau theorem,
this matter is allowed to deform the bulk geometry so that the
bulk is \emph{not} an Einstein manifold.

In the case of the meron, one begins with the compactification of
eleven-dimensional supergravity on S$^7$. This has a consistent
truncation to a theory with an SU(2) gauge field and a graviton.
Maldacena and Maoz consider a gauge field with a Lagrangian
density
\begin{equation}\label{eq:G}
\alpha\sqrt{g}\;F^a_{\mu\nu}F^{a\mu\nu},
\end{equation}
where $\alpha$ is a non-negative constant. Adding this to the
Einstein-Hilbert Lagrangian density and including a negative
cosmological constant $-$3/L\2 [since we want an ``AdS\4-like
cosmology", that is, one which reverts to AdS\4 if $\alpha$ is
zero], Maldacena and Maoz solve the resulting field equations and
obtain the Euclidean metric
\begin{equation}\label{eq:GG}
g^+_{MM} = \; dt^2 +
\textup{L}^2[(\alpha+{{1}\over{4}})^{1/2}\textup{cosh}({{2t}
\over{\textup{L}}})-{{1}\over{2}}]\times[d\chi^2 +
\textup{sin}^2(\chi)\{d\theta^2 +
\textup{sin}^2(\theta)d\phi^2\}],
\end{equation}
defined on a manifold with universal cover of the form
$\bbr\;\times\;$S\3. The Lorentzian version,
\begin{equation}\label{eq:GGG}
g^-_{MM} = \;-\; dt^2 +
\textup{L}^2[(\alpha+{{1}\over{4}})^{1/2}\textup{cos}({{2t}
\over{\textup{L}}})-{{1}\over{2}}]\times[d\chi^2 +
\textup{sin}^2(\chi)\{d\theta^2 +
\textup{sin}^2(\theta)d\phi^2\}],
\end{equation}
is indeed a Bang/Crunch cosmology with locally spherical spatial
sections: computing the invariant
\begin{equation}\label{eq:GGGG}
[\textup{R}^{\mu\nu} \;+\;{{3}\over
{L^2}}g^{\mu\nu}\;]\;[\textup{R}_{\mu\nu} \;+\;{{3}\over
{L^2}}g_{\mu\nu}\;]\;=\; {{12\alpha^2/L^4}\over
{[(\alpha+{{1}\over{4}})^{1/2}\textup{cos}({{2t}
\over{L}})-{{1}\over{2}}]^4}},
\end{equation}
we see that the singularities in $g^-_{MM}$ at times t$_B$, t$_C$,
given by
\begin{equation}\label{eq:GGGGG}
-\;t_B\;=\;t_C\;=\;L\;\textup{cos}^{-1}(1/\sqrt{1+4\alpha}),
\end{equation}
are genuine curvature singularities at a Bang and a Crunch
respectively. Notice that the total proper lifetime of the
Maldacena-Maoz universe is
$2L\;\textup{cos}^{-1}(1/\sqrt{1+4\alpha})$, which becomes
\emph{shorter} as $\alpha$ is reduced. One can imagine that the
Yang-Mills fields are ``holding apart" the Bang and the Crunch.
The full extent of conformal time for this spacetime,
\begin{equation}\label{eq:GGGGGG}
\int_{t_B}^{t_C}{{dt}\over
{\sqrt{(\alpha+{{1}\over{4}})^{1/2}\textup{cos}({{2t}
\over{L}})-{{1}\over{2}}}}}
\end{equation}
is always less than $\pi$L; for example, it is about
2.384$\times$L if $\alpha$ = 0.75. As the covering spacetime is
conformal to part of the Einstein static universe
\cite{kn:hawking}, the precise shape of the Penrose diagram will
depend on the choice of topology for the spatial sections: if they
have the topology of S\3, then the diagram will be a rectangle
which is wider than it is high, meaning that [unlike, for example,
in a matter-dominated FRW cosmology with spherical spatial
sections], the particle horizons never disappear, even during the
contraction phase. On the other hand, if the spatial sections are
[for example] copies of $\bbr$P\3 [see \cite{kn:rp3}] then this
will not be so; this is of course a possibility consistent with
the metric (\ref{eq:GGG}). The Penrose diagram will have the form
shown in Figure (1) if we choose $\alpha$ = 0.75 and $\bbr$P\3
spatial sections. [The diamonds on the right indicate that points
there represent copies of $\bbr$P\2 rather than two-spheres.]
\begin{figure}[!h]
\centering
\includegraphics[width=0.3\textwidth]{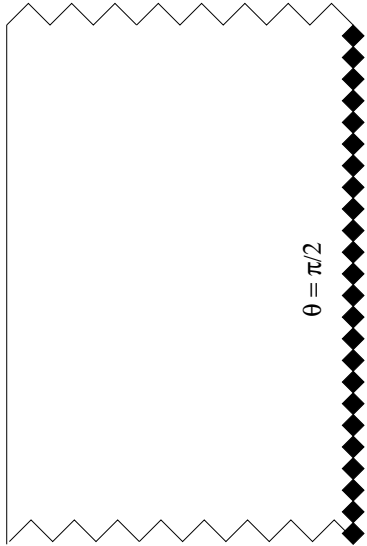}
\caption{Penrose diagram of MM spacetime, $\alpha$ = 0.75,
$\bbr$P\3 sections}
\end{figure}
By inspecting this diagram we see that there is no way that the
spacetime can be extended. Therefore the coordinates used in
(\ref{eq:GGG}) are global coordinates, and it follows that there
is no timelike Killing vector here --- the geometry is genuinely
time-dependent. [By contrast, a Penrose diagram of the region of
AdS\4 covered by the coordinates used in (\ref{eq:A}) would
immediately reveal that this region can be extended, and that
there is a timelike Killing vector once this extension is
performed.]

The reader has no doubt observed that the Maldacena-Maoz spacetime
is very different indeed to the AdS\4 from which it originates: it
is singular, globally hyperbolic, and has a time-dependent
geometry, while AdS\4 has none of these properties. In fact, the
Einstein equations for FRW spacetimes imply that, because the
spatial sections here are positively curved, the energy density of
the Yang-Mills fields must be greater [at all times] in absolute
value than the energy density contributed by the cosmological
constant. Thus, the Maldacena-Maoz spacetime is not a ``small
perturbation" of AdS\4. Nevertheless, the Euclidean version of
this spacetime does have almost the same asymptotic geometry as
the Euclidean version of AdS\4 --- indeed, at sufficiently large
distances, the only real difference is precisely the fact that the
Euclidean Maldacena-Maoz spacetime has two asymptotic regions.
This is the key virtue of the Maldacena-Maoz proposal: even though
Bang/Crunch cosmologies are vastly different from AdS\4, their
Euclidean versions are sufficiently similar as to warrant hope
that a holographic description is possible.

Returning to the Euclidean version given by equation
(\ref{eq:GG}), note that if the meron is turned off by setting
$\alpha$ = 0, then a simple calculation shows that this is just
the standard metric for four-dimensional Euclidean anti-de Sitter
space. That is, we obtain ordinary hyperbolic space H$^4$, with
the metric expressed in polar coordinates, if the sections are
copies of S\3. [If the sections are copies of $\bbr$P\3, then we
obtain an orbifold of H$^4$, but this orbifolding only happens if
$\alpha$ is \emph{exactly} zero.] If $\alpha$ does not vanish,
then the locally spherical sections do not shrink down to zero
size at t = 0 as they do in H$^4$. Instead they open up again to a
second region like the first. [Geometrically this is rather like a
smooth version of the wormhole constructed in \cite{kn:adsworm}.
The physical difference is that that wormhole required exotic
matter in the form of a negative-tension brane at the throat, the
bulk being otherwise matter-free.] The conformal boundary
consists, again in the most obvious interpretation, of two copies
of some three-manifold with the local geometry of S\3, which of
course has a metric of positive scalar curvature. The bulk is
geodesically complete, but it is not an Einstein manifold, so the
simplest version of the Witten-Yau theorem, which assumes that the
bulk is an Einstein manifold, does not apply here.

As we saw, however, this alone is not enough: the Witten-Yau
theorem can handle \emph{some} non-Einstein manifolds. Indeed, the
Witten-Yau stipulation that the bulk matter should increase the
Ricci curvature [that is, it should make the Ricci
  curvature less negative than it is in anti-de Sitter spacetime]
\emph{seems} very reasonable --- it looks very much like a
Euclidean version of the Strong Energy Condition [which just
requires that a given form of matter should have an energy density
making a non-negative contribution to the Ricci curvature]. In
fact, the relevant Ricci component for $g^-_{MM}$ is
\begin{equation}\label{eq:PPPP}
\textup{R}_{00}(g^-_{MM}) \;=\; \;{{3}\over {L^2}}\; + \;
{{3\alpha/L^2}\over
{[(\alpha+{{1}\over{4}})^{1/2}\textup{cos}({{2t}
\over{L}})-{{1}\over{2}}]^2}},
\end{equation}
and we see explicitly that the Yang-Mills field in the Lorentzian
Maldacena-Maoz cosmology does indeed make a positive contribution
to the Ricci curvature. [Actually, in agreement with the general
discussion above, a simple calculation shows that, at all times,
its contribution is larger than that of the cosmological constant
itself.] In fact, Yang-Mills fields always satisfy the Strong
Energy Condition. One might have expected the Witten-Yau theorem
to forbid a double boundary here; but evidently it does not. What
is going wrong?

There are actually \emph{two} things ``going wrong" here, and it
is important to be clear about this, because one of the problems
is more important than the other. Let us explain.

\addtocounter{section}{1}
\section*{4. Escaping the Menace of the WY Theorem, Part 1}

The first reason that the Witten-Yau theorem [even in the version
which does not require the bulk to be an Einstein manifold] does
not apply to the Maldacena-Maoz manifold is that matter which
satisfies the Strong Energy Condition \emph{does not necessarily
cause the Ricci curvature to increase in all directions of the
Euclidean version. } The reason for this can be seen in the
following elementary way.

Consider a Euclidean field theory in (n+1) dimensions with an
energy-momentum tensor T$_{\mu\nu}$. Diagonalising, we can express
T with respect to an orthonormal basis as
\begin{equation}\label{eq:J}
\textup{T}_{\mu\nu} = {\hbox{diag}} (p_0, p_1, p_2, ..., p_n),
\end{equation}
where the p$_i$ are the eigenvalues. [Henceforth, Greek letters
are spacetime indices; all other indices are just labels, as in
(\ref{eq:E}) and (\ref{eq:F}) above.] The Einstein equation [with
cosmological constant] gives us Ricci eigenvalues
\begin{equation}\label{eq:K}
\textup{Ric}_{(i)} = -{n\over L^2} + p_i - {1\over n-1} \;
\sum^n_{j=0} \; p_j.
\end{equation}
From this we immediately see that the condition for the
introduction of matter into AdS$_{n+1}$ to increase the Ricci
eigenvalues is just
\begin{equation}\label{eq:L}
p_i \; \geq \; {1\over n-1} \; \sum^n_{j=0} \; p_j \;\;\forall i,
\end{equation}
or
\begin{equation}\label{eq:M}
p_i \; \geq \; {1\over n-2} \; \sum^n_{j \neq i} \; p_j
\;\;\forall i.
\end{equation}
To see what this means, consider the five-dimensional case [n=4],
so that we have
\begin{equation}\label{eq:N}
p_0 \; \geq \; {{1}\over{2}}(p_1 + p_2 + p_3 + p_4),
\end{equation}
and likewise
\begin{equation}\label{eq:O}
p_1 \; \geq \; {{1}\over{2}}(p_0 + p_2 + p_3 + p_4).
\end{equation}
Combining these we have
\begin{equation}\label{eq:P}
p_0 \; \geq \;  p_2 + p_3 + p_4,
\end{equation}
and similarly with the roles of p$_0$ and p$_2$ reversed, whence
it follows that p$_3$ + p$_4$ must be non-positive, and of course
the same applies to any distinct pair of eigenvalues. This means
that of the p$_i$, at most one can be positive. Clearly a similar
argument works in all dimensions. But this is an unreasonably
restrictive requirement; for example, it is easy to construct
Yang-Mills configurations, in any Euclidean dimension, such that
the energy-momentum tensor has more than one positive eigenvalue
--- see below for an example. The Witten-Yau inequalities
(\ref{eq:F}) therefore do not apply to such fields, despite the
fact that the Lorentzian versions satisfy the Strong Energy
Condition. In short, the inequalities (\ref{eq:F}) cannot in
general be motivated by imposing this energy condition; in fact,
they apparently require that the SEC be violated.

In the specific case of the Euclidean metric (\ref{eq:GG}) studied
by Maldacena and Maoz, the eigenvalue of the Ricci tensor
corresponding to the coordinate t may be computed as
\begin{equation}\label{eq:PP}
\textup{Ric}_{(0)} \;=\; -\;{{3}\over {L^2}}\; - \;
{{3\alpha/L^2}\over
{[(\alpha+{{1}\over{4}})^{1/2}\textup{cosh}({{2t}
\over{L}})-{{1}\over{2}}]^2}},
\end{equation}
and since the Yang-Mills parameter $\alpha$ is non-negative, we
see at once that this particular eigenvalue is in fact
\emph{decreased} by the presence of the matter. [The other three
are increased; since the Yang-Mills energy-momentum tensor is
traceless in four dimensions, the Ricci tensor is proportional to
this energy-momentum tensor, and so these three positive
contributions mean that there are three positive eigenvalues of
the energy-momentum tensor, again showing, in view of our earlier
more general discussion, that the inequalities (\ref{eq:F}) are
not satisfied here.]

Thus we see explicitly that the Maldacena-Maoz manifold violates
the Witten-Yau inequalities. To see this in a more dramatic way,
notice that at the wormhole throat [t = 0] we have from
(\ref{eq:PP}) that
\begin{equation}\label{eq:PPP}
\textup{Ric}_{(0)}(\textup{Throat})\; + \;{{3}\over {L^2}}\; =
\;{{- 3 [(\alpha+{{1}\over{4}})^{1/2}\;+\; {{1}\over{2}}]^2}
\over{\alpha L^2}},
\end{equation}
from which we derive the interesting fact that if $\alpha$ is very
small, so that the geometry is almost indistinguishable from that
of Euclidean AdS\4 except very near to the throat, then the extent
of the violation of the Witten-Yau inequalities is very
\emph{large}, not small: the inequalities are \emph{not} close to
being satisfied [near the throat] in this case.

For Yang-Mills fields in four dimensions, the situation we have
been discussing is in fact generic: the ``energy-momentum tensor"
in that dimension must be traceless, and so the same is true of
its contribution to the Ricci curvature. Since the sum is zero,
there must, in any non-trivial configuration, be both positive and
negative contributions, and so it is clear that the inequalities
(\ref{eq:F}) cannot possibly be satisfied here.  Thus the
Witten-Yau theorem does not apply to \emph{any} non-trivial
Yang-Mills configuration in a four-dimensional asymptotically
anti-de Sitter spacetime. [In \cite{kn:adsworm}, the Witten-Yau
theorem was avoided in a different way: the conditions
(\ref{eq:F}) are satisfied everywhere except at a negative-tension
brane, but, because of the presence of the brane, the bulk is not
geodesically complete, and this too renders the Witten-Yau theorem
inapplicable.]

To summarize: the mere fact that matter has been inserted into
AdS\4 does not explain the ability of the Maldacena-Maoz
cosmologies to have disconnected Euclidean boundaries with
positive scalar curvature: the presence of matter is necessary but
not sufficient. The simplest reason for this ability --- though,
as we shall see, even this is far from the full explanation --- is
just that, contrary to intuition, the well-behaved matter used in
\cite{kn:maldacena} violates the Witten-Yau inequalities.

 The question now is this: suppose that we consider other
forms of bulk matter. Do these, too, naturally violate the
inequalities (\ref{eq:F}), or is this a specific property of
Yang-Mills fields? If the latter is the case, then one might try
to ascribe the disconnectedness of the boundary to a
cosmologically unrealistic choice of bulk matter.

In fact, the true situation is much more complex. To understand
why, consider scalar matter instead of Yang-Mills fields. Here,
for a scalar field $\varphi$ with the usual kinetic term and with
potential V($\varphi$), the energy-momentum tensor is
\begin{equation}\label{eq:Q}
\textup{T}_{\mu\nu} = \partial_\mu \varphi \; \partial_\nu\varphi
- {1\over 2} g_{\mu\nu} (\partial_\alpha\varphi \;
\partial^\alpha\varphi) - g_{\mu\nu}\; \textup{V}(\varphi).
\end{equation}
Inserting such matter into AdS\4, we have
\begin{equation}\label{eq:R}
\textup{R}_{\mu\nu} \;=\;-\; {3\over \textup{L}^2}\; g_{\mu\nu}
\;+\;
\partial_\mu \varphi \;
\partial_\nu \varphi + g_{\mu\nu} \; \textup{V}(\varphi).
\end{equation}
Now if t is the proper time of a Lorentzian FRW spacetime obtained
in this way, we find that the corresponding component of the Ricci
tensor is
\begin{equation}\label{eq:RR}
\textup{R}_{00}\;=\;{3\over \textup{L}^2}\; +\; \dot{\varphi}^2
\;-\;\textup{V}(\varphi).
\end{equation}
We see that a {\it positive} V($\varphi$) will {\it reduce}
R$_{00}$; this \emph{may or may not} lead --- leaving aside the
cosmological constant --- to violations of the Strong Energy
Condition. It will do so if the rate of evolution of the scalar
field becomes sufficiently small; this often happens in a de
Sitter-like spacetime, but \emph{not necessarily} in an anti-de
Sitter-like spacetime, where the Bang and the Crunch may keep the
kinetic term sufficiently large so that the SEC is always
satisfied. [Actually it does not necessarily occur even in the de
Sitter-like case, as for example in some kinds of ``eternal"
quintessence.] In short, the status of the SEC in scalar field
physics is ambiguous: it is violated in some circumstances but not
in others.

Now let us consider the Euclidean case. There is a very
interesting ambiguity as to how a scalar field should be
``Euclideanized", particularly in the context of Euclidean
wormholes. It was argued by Giddings and Strominger \cite{kn:GS}
that an ordinary scalar field cannot generate topologically
non-trivial Euclidean configurations, such as wormholes, in the
asymptotically flat context; instead they considered a
\emph{massless axion}, which, when expressed in terms of a locally
equivalent single-component field, has a Euclidean energy-momentum
tensor of the opposite sign to that of the usual massless scalar
Euclidean energy-momentum tensor. As is emphasised in
\cite{kn:GS}, this sign reversal is the key property needed to
allow topologically non-trivial asymptotically flat
configurations, for it allows the axion to make a negative
contribution to the Euclidean Ricci curvature. The theorem cited
by Giddings and Strominger \cite{kn:cheeger} does not apply here
[among other things, the spaces we shall consider do not have any
boundary of topology S\3, and, more importantly, they are not
asymptotically flat], and in fact the situation considered by
Witten and Yau is considerably more delicate than the one covered
by that theorem. However, the fact that the Witten-Yau theorem
attempts to rule out wormholes by assuming that matter sources
make a positive contribution to the Euclidean Ricci tensor,
combined with the Giddings-Strominger observation that axions do
\emph{not} make such contributions in the case of the wormholes
they considered, suggests strongly that some kind of axionic
matter is relevant to the kind of cosmological wormholes proposed
by Maldacena and Maoz.

We are thus led to consider quintessence-like cosmological matter
fields, where the usual quintessence field is replaced by a
``generalized axion". The requirements of cosmology will force us
to consider unconventional potentials --- which is what we shall
mean by ``generalized". Lorentzian axions with conventional
potentials have in fact been discussed as dark energy candidates
[see for example \cite{kn:choi}\cite{kn:kim}], but they are not
suitable for our purposes, since the effective axion potential is
bounded, which does not seem natural when approaching a Bang or a
Crunch.

The matter field we shall consider in the next section shares with
the Giddings-Strominger axion the ability [for a certain parameter
range] to make a negative contribution to the Euclidean Ricci
curvature; it is in fact a generalized axion in the above sense.
Let us therefore consider the Euclidean geometry corresponding to
a generalized axion; it is convenient to do this by taking
equation (\ref{eq:R}), reversing the signature, and complexifying
the scalar field. Then we have Euclidean Ricci eigenvalues given
by
\begin{equation}\label{eq:S}
\textup{Ric}_{(0)}\;=\;-\;{3\over \textup{L}^2}\;
-\;\dot{\varphi}^2 \;+\;\textup{V}_e(\varphi)
\end{equation}
and
\begin{equation}\label{eq:SS}
\textup{Ric}_{(1)}\;=\;\textup{Ric}_{(2)}\;=\;\textup{Ric}_{(3)}\;=
\;-\;{3\over \textup{L}^2}\;+\;\textup{V}_e(\varphi),
\end{equation}
where V$_e$($\varphi$) is the Euclidean potential.

Obviously the last three eigenvalues will always exceed $-$3/L\2
if V$_e$($\varphi$) is positive [which is certainly the case for
the potential we shall consider below]. Even Ric$_{(0)}$ will do
so if V$_e$($\varphi$) is positive and sufficiently large; on the
other hand, with this Euclideanization, it can also fall below
$-$3/L\2 if the Euclidean potential is too small. Thus, in sharp
contrast to the Yang-Mills case, it is \emph{possible} to satisfy
the Witten-Yau inequalities with generalized axion matter
--- though it is also possible to violate them, in the Giddings-Strominger manner. However, the WY inequalities can
only be satisfied with the aid of potentials of a kind which tend
to \emph{violate} the Strong Energy Condition in the Lorentzian
case, that is, with potentials which are positive.

It is of course easy to arrange scalar potentials which violate
the SEC: that is the point of quintessence. The Euclidean version
can then satisfy the Witten-Yau inequalities, and so --- one would
think --- disconnected boundaries should be forbidden if we use
such matter in a FRW cosmology with positively curved or flat
spatial sections. [Recall that Cai and Galloway \cite{kn:galloway}
extended the Witten-Yau theorem to the [scalar] flat case, and
that in the cosmological application the spatial sections have the
same geometry as the boundary components of the Euclidean
version.] If this were the case, then one could argue that
disconnected boundaries are not a matter of concern, \emph{since
they apparently cannot arise in cosmological models with realistic
[SEC-violating] matter content}.

Puzzling as it seems, this is wrong, however: evidently we have
not yet fully understood the ability of the Euclidean
Maldacena-Maoz spaces to have disconnected boundaries. To see
this, we shall explicitly construct a family of AdS-like
quintessence cosmologies such that the Euclidean boundary is
disconnected even though the Witten-Yau inequalities are
satisfied. These cosmologies have some independent interest.

\addtocounter{section}{1}
\section*{5. Quintessence Instead of Yang-Mills}
As we saw, Maldacena and Maoz construct their spacetimes by
introducing Yang-Mills matter into anti-de Sitter spacetime. Here
we shall follow their example, but with a different choice of bulk
matter. This bulk matter will be quintessence \cite{kn:ratra} with
a particular choice of potential. Ultimately, as discussed in the
previous section, we might wish to regard it as ``generalized
axion" matter, but for simplicity we shall at first present it as
an ordinary quintessence field. The entire discussion will be
Lorentzian until further notice.

A popular choice of quintessence potential is obtained by
combining exponentials of the scalar field $\varphi$, since these
can be motivated both by fundamental physics and by astrophysical
arguments. For example, such potentials arise naturally in
supergravity
--- see the recent discussion \cite{kn:gutperle} in connection with Cosmic
Censorship --- and also M-theory
\cite{kn:townsend}\cite{kn:ish1}\cite{kn:jarv}. On the
astrophysical side, in \cite{kn:cardenas} and \cite{kn:rubano}
potentials similar to the one below are used, and the
observational consequences are explored; see \cite{kn:kehagias}
for an extensive list of references on the use of exponential
potentials in cosmology.

Our objective here is not to find a model which is completely
realistic; we merely seek an accelerating analogue of the
spacetimes considered by Maldacena and Maoz. For our purposes, it
is essential to obtain \emph{exact} solutions for the metric, so
that the geometric properties of the spacetime can be analysed
precisely; thus we do not allow any other form of matter apart
from the AdS\4 cosmological constant and the quintessence field.
We choose a very simple quintessence potential which is
approximately exponential for both large negative and large
positive values of $\varphi$. [It is also well-behaved under
complexification of $\varphi$, that is, it remains real and
positive; this is important for the axionic interpretation when we
eventually turn to the Euclidean version.] As usual, the potential
has two free parameters, one scaling $\varphi$ and one scaling the
overall potential. For later convenience we choose these two
parameters, $\varpi$ and $\xi$, so that the potential has the
following form:
\begin{equation}\label{eq:H}
\textup{V}(\varphi) =  {{3
\;-\;\varpi^{-1}}\over{\xi^2}}\textup{cosh}^2(\sqrt{{{4\pi}\over{\varpi}}}\;\;\varphi).
\end{equation}
We shall always take $\varpi$ to satisfy
\begin{equation}\label{eq:HHH}
\varpi\;\ > \;{{1}\over{3}},
\end{equation}
so that V($\varphi$) is \emph{strictly positive}. The
parametrization is chosen so that, in the appropriate limit
[$\varpi \rightarrow \infty$], the potential tends to a constant;
so this is the limit in which quintessence becomes a positive
cosmological constant. Differentiating V($\varphi$) we obtain
after some elementary algebra the following identity between
V($\varphi$) and its first derivative:
\begin{equation}\label{eq:HH}
({{\textup{dV}}\over{\textup{d}\varphi}})^2 \; - \; {{16\pi}\over
{\varpi}}\;\textup{V}^2 \; + \; {{48\pi}\over {\varpi \xi^2}}(1 \;
- \; {{1}\over {3\varpi}})\;\textup{V} \; = \;  0,
\end{equation}
a relation which we shall use below.

We now propose to introduce this kind of matter into an AdS\4
background with cosmological constant $-$3/L\2. We shall search
for Lorentzian FRW solutions of the Einstein equations with flat
but compact spatial sections, so that the metric will have the
general form
\begin{equation}\label{eq:T}
g^- = -dt^2 + A^2\;a(t)^2[d\theta_1^2 + d\theta_2^2 +
d\theta_3^2],
\end{equation}
as in equation (\ref{eq:B}) above, so that A measures the
circumferences of the torus, and $a$(t) is the scale factor. [As
before, we want tori here because we wish ultimately to consider
an AdS/CFT kind of scenario for cosmology, and it is preferable
for that purpose that the boundary should be compact. Note however
that toral [or, more generally, compact flat] spatial sections are
natural in many cosmological models, as for example those
considered in most brane gas models: see
\cite{kn:brand}\cite{kn:watson} and their references.]

With the usual kinetic term, the density and pressure
corresponding to $\varphi$ are
\begin{equation}\label{eq:I}
\rho\;=\; {{1}\over{2}}\;\dot{\varphi}^2 \;+\;\textup{V}(\varphi)
\end{equation}
and
\begin{equation}\label{eq:U}
\textup{p}\;=\; {{1}\over{2}}\;\dot{\varphi}^2
\;-\;\textup{V}(\varphi)
\end{equation}
respectively, and so the Einstein equation for FRW spacetimes with
flat spatial sections becomes
\begin{equation}\label{eq:V}
({{\dot{a}\over{a}}})^2 \;=\;
{{8\pi}\over{3}}\;[\;{{1}\over{2}}\;\dot{\varphi}^2
\;+\;\textup{V}(\varphi) \;-\;{{3}\over{8\pi L^2}}].
\end{equation}
The field equation for $\varphi$ is
\begin{equation}\label{eq:W}
\ddot{\varphi}\;+\;3\;{{\dot{a}\over{a}}}\;\dot{\varphi}\;+\;{{\textup{dV}}\over{\textup{d}\varphi}}\;=\;0.
\end{equation}
This may be usefully re-written as
\begin{equation}\label{eq:X}
\dot{\varphi}^2({{\dot{a}}\over{a}})^2\;=\;{{1}\over{9}}\;[\;{{\textup{d}}\over{\textup{d}\varphi}}\;
[\;{{1}\over{2}}\;\dot{\varphi}^2\;+\;\textup{V}(\varphi)\;]\;]^2.
\end{equation}
Substituting this into the Einstein equation (\ref{eq:V}) we have
\begin{equation}\label{eq:Y}
(\;{{\textup{d}}\over{\textup{d}\varphi}}\;
[\;{{1}\over{2}}\;\dot{\varphi}^2\;+\;\textup{V}(\varphi)\;]\;)^2\;=\;24\pi\;\dot{\varphi}^2\;[\;{{1}\over{2}}\;\dot{\varphi}^2
\;+\;\textup{V}(\varphi) \;-\;{{3}\over{8\pi L^2}}].
\end{equation}
After eliminating the derivative of V($\varphi$) using
(\ref{eq:HH}), we can regard this as a relation between $\varphi$
and $\dot{\varphi}^2$ only, and it can be solved for the latter in
terms of the former; inverting V($\varphi$) we can in principle
solve for $\dot{\varphi}^2$ in terms of V($\varphi$). We can spare
ourselves that onerous task by noticing the structural similarity
of (\ref{eq:Y}) with the identity (\ref{eq:HH}), which suggests
the simple ansatz
\begin{equation}\label{eq:Z}
\dot{\varphi}^2\;=\;\textup{K}\;\textup{V}(\varphi),
\end{equation}
where K is a constant to be determined by comparing (\ref{eq:Y})
with (\ref{eq:HH}). We find that indeed this solves (\ref{eq:Y})
\emph{provided} that $\xi$ is equated to $\sqrt{8\pi}\;$L, that
(\ref{eq:HHH}) holds, and that K is chosen to be
2/(3$\varpi\;-\;1$). [We stress that this procedure only works if
we strictly enforce the inequality (\ref{eq:HHH}); this has the
unfortunate consequence that our subsequent formulae will not
reflect the fact, visible in (\ref{eq:H}), that we recover AdS\4
when $\varpi$ = 1/3, that is, when the scalar field is switched
off.]

Thus by fixing the value of $\xi$ [thereby reducing to a
one-parameter subset of solutions], we can solve for
$\dot{\varphi}^2$ in terms of V($\varphi$), obtaining
\begin{equation}\label{eq:ZZ}
\dot{\varphi}^2\;=\;{{2}\over{3\varpi\;-\;1}}\;\textup{V}(\varphi).
\end{equation}
This no longer involves the scale factor and so it can be solved
directly:
\begin{equation}\label{eq:ZZZ}
\varphi\;=\;\pm\sqrt{{{\varpi}\over{4\pi}}}\;\textup{cosh}^{-1}(\textup{sec}({{t}\over{\varpi\textup{L}}})),
\end{equation}
where the sign agrees with that of t [so that $\varphi$(t) is
smooth --- note that $\dot{\varphi}$ is never zero], and where
cosh$^{-1}$ is defined to be non-negative.

Equation (\ref{eq:ZZ}) can also be used to find the scale factor.
Substituting it into equations (\ref{eq:I}) and (\ref{eq:U}) we
find expressions for the quintessence density and pressure in
terms of the potential,
\begin{equation}\label{eq:AA}
\rho \;=\; {{3\varpi}\over {3\varpi\;-\;1}}\;\textup{V}(\varphi)
\end{equation}
and
\begin{equation}\label{eq:AB}
\textup{p} \;=\; {{2\;-\;3\varpi}\over
{3\varpi\;-\;1}}\;\textup{V}(\varphi).
\end{equation}

Now the fact that the overall energy-momentum tensor is
divergenceless gives us the standard relation
\begin{equation}\label{eq:AC}
\dot{\rho}\;+\;3\;{{\dot{a}}\over
{a}}\;(\rho\;+\;\textup{p})\;=\;0;
\end{equation}
eliminating the time derivatives and using (\ref{eq:AA}) and
(\ref{eq:AB}) we obtain from this
\begin{equation}\label{eq:AD}
{{\textup{d}ln(\rho)\over {\textup{d}ln(a)}}}\;=\;-\;2/\varpi.
\end{equation}
This gives us
\begin{equation}\label{eq:AE}
\rho\;=\;C\;a^{(-\;2/\varpi)},
\end{equation}
where C is a constant which we can fix as follows. We are
interested in obtaining Bang/Crunch cosmologies. This means that
there must be a time when the scale factor reaches a maximum and
so has zero time derivative. From the Einstein equation
(\ref{eq:V}) we see that, at this time, the total density
[negative anti-de Sitter density plus positive quintessence
density] must vanish. Because the spatial sections are flat, their
intrinsic scale is not fixed by the other parameters so we are
free to require that $a$(t) should be equal to unity at this time.
[This simply means that we are defining the parameter A in
equation (\ref{eq:T}) to be such that the circumference of the
\emph{maximal} toral spatial section is 2$\pi$A.] Thus $\rho$ must
equal 3/8$\pi$L\2 when $a$ = 1, and this fixes C at 3/8$\pi$L\2.
Substituting (\ref{eq:AE}) with this value of C back into the
Einstein equation (\ref{eq:V}), we obtain a differential equation
for $a$(t):
\begin{equation}\label{eq:AF}
({{\dot{a}\over{a}}})^2 \;=\;{{1}\over
{L^2}}[\;a^{(-\;2/\varpi)}\;-\;1].
\end{equation}
This equation has an \emph{exact} solution: defining t = 0 to be
the time of maximum expansion, we have
\begin{equation}\label{eq:AG}
a(t)\;=\;\textup{cos}^{\varpi}({{t}\over {\varpi L}}),
\end{equation}
and so we arrive finally at a family of  ``quintessential
Maldacena-Maoz" Lorentzian metrics, parametrised by A and
$\varpi$, given by the very simple metric
\begin{equation}\label{eq:AH}
g^-(\varpi,A) = -dt^2 + A^2\;\textup{cos}^{2\varpi}({{t}\over
{\varpi \textup{L}}})\;[d\theta_1^2 + d\theta_2^2 + d\theta_3^2].
\end{equation}
Note that the metric $g^-(1,A)$, given by equation (\ref{eq:B}),
which we obtained simply by replacing the negatively curved
spatial sections of AdS\4 by flat tori, is indeed the special case
$\varpi$ = 1. The more general metric $g^-(\varpi,A)$ may be
thought of in the same way: replace the negatively curved spatial
sections of AdS\4 by flat spaces and replace cos$^2$(t/L) with
cos$^{2\varpi}$(t/$\varpi$L).

The metrics $g^-(\varpi,A)$ apparently represent universes which
have a Big Bang at t = $-\pi\varpi$L/2, expand to a maximum size
at t = 0, and then collapse to a Big Crunch at $+\pi\varpi$L/2.
However, we know from the AdS\4 example that such appearances can
be deceptive, so let us verify that these spacetimes are indeed
singular.

The scalar curvature is given by
\begin{equation}\label{eq:AI}
\textup{R}(g^-(\varpi,A))\;=\;-\;{{12}\over{\textup{L}^2}}\;+\;{{6}\over{\textup{L}^2}}\;[2\;-\;{{1}\over
{\varpi}}]\;\textup{sec}^2({{t}\over {\varpi \textup{L}}}),
\end{equation}
which immediately shows that the metrics for different values of
$\varpi$ are distinct and that all of these metrics are singular
at t = $\pm \pi\varpi$L/2, with the possible exception of
$g^-(1/2,A)$, which can be shown to be singular in other ways.
[Consider again equation (\ref{eq:B}): the metric $g^-(1,A)$
differs from that of AdS\4 only in that the spatial sections have
been flattened. We now see that this flattening causes the
spacetime to become singular, as claimed.]

We see that these are indeed Bang/Crunch cosmologies with a total
lifetime given by $\pi\varpi$L. This number must of course be
large, significantly larger than the current age of the Universe.
As we shall soon see, observational evidence can in principle fix
$\varpi$ and L separately, and such data as we have suggest that
$\varpi$ must be large; for definiteness we shall assume that L is
roughly equal to the age of the Universe, so that the factor
$\pi\varpi$ is responsible for stretching the time scale. From the
AdS/CFT point of view \cite{kn:oz}, the value of L in pure AdS\4
is related to the strength of the coupling of the boundary field
theory. Here we do not have pure AdS\4, but we see from
(\ref{eq:AI}) that L continues to set the overall scale of the
curvature; so we shall continue to assume that if these spacetimes
have a holographic interpretation, then L is still a rough measure
of the strength of the coupling in the dual field theory. By
taking L to be of a typical cosmological size, we are implicitly
assuming that the field theory is strongly coupled. We shall see
that large values of $\varpi$ lead to particularly interesting
models.

The time-time component of the Ricci tensor is given by
\begin{equation}\label{eq:AII}
\textup{R}_{00}(g^-(\varpi,A))\;=\;{{3}\over{\textup{L}^2}}\;-\;{{3}\over{\textup{L}^2}}\;[1\;-\;{{1}\over
{\varpi}}]\;\textup{sec}^2({{t}\over {\varpi \textup{L}}}).
\end{equation}
Clearly quintessence always makes a non-negative contribution to
this component of the Ricci tensor, that is, the Strong Energy
Condition is satisfied at all times, if and only if $\varpi \;
\leq \;$ 1. For these values of $\varpi$, the cosmology is much
like the traditional dust FRW model with density greater than the
critical density: the Universe decelerates at all times from a
Bang to a Crunch.

We shall be more interested in values of $\varpi$ significantly
greater than unity. For these, we see that the Strong Energy
Condition is only satisfied for an interval of time around t = 0,
namely the interval
\begin{equation}\label{eq:AIII}
|\;\textup{t}|\; \leq \; \varpi
\textup{L}\;\textup{cos}^{-1}\sqrt{1\;-\;{{1}\over {\varpi}}}\;.
\end{equation}
As a fraction of the total duration of the Universe, the length of
this interval is
\begin{equation}\label{eq:AJ}
\Delta \; = \; {{2}\over
{\pi}}\;\textup{cos}^{-1}\sqrt{1\;-\;{{1}\over {\varpi}}}\;,
\end{equation}
which is a decreasing function of $\varpi$; for very large
$\varpi$, the era in which the SEC holds is a very small fraction
of the total duration of the universe. Notice that this is
independent of the value of L.

In order to show how the values of $\varpi$ and L can in principle
be fixed in terms of cosmological observations, we begin by noting
that the current value of the Hubble constant in these cosmologies
is given according to (\ref{eq:AF}) and (\ref{eq:AG}), by
\begin{equation}\label{eq:AHH}
\textup{H}_0\;=\;{{1}\over{\textup{L}}}\;\textup{cot}({{\textup{T}}\over{\varpi
\textup{L}}}),
\end{equation}
where T is the current age of the Universe, that is, measured from
t = $-\pi\varpi$L/2; regarding T and H$_0$ as known from
observations, we have one relation between $\varpi$ and L.

To obtain another, observe that the equation-of-state parameter w
for these cosmologies [that is, the ratio of the total pressure to
the total density] is given, using equations (\ref{eq:AA}) and
(\ref{eq:AB}), by
\begin{equation}\label{eq:AK}
\textup{w} \; = \; {{{{3}\over {8\pi
L^2}}\;+\;{{2\;-\;3\varpi}\over {3\varpi}}\;\rho}\over
{-\;{{3}\over {8\pi L^2}}\;+\;\rho}}.
\end{equation}
Using equation (\ref{eq:AE}), with C fixed at 3/8$\pi$L\2, we can
write this simply as
\begin{equation}\label{eq:AL}
\textup{w} \;= \; -\;1\;+\;{{2}\over
{3\varpi}}\;\textup{cosec}^2({{t}\over {\varpi \textup{L}}});
\end{equation}
as in pure quintessence scenarios, the theory cannot tolerate
values of w below $-$1, and would be ruled out by firm evidence in
favour of such values. Evaluating w at the present time, we have
\begin{equation}\label{eq:ALL}
\textup{w}_0 \;= \; -\;1\;+\;{{2}\over
{3\varpi}}\;\textup{sec}^2({{\textup{T}}\over {\varpi
\textup{L}}}),
\end{equation}
where again T is the current age of the Universe. In principle,
w$_0$ can be fixed by observation: in practice this can only be
done rather roughly, but there is evidence \cite{kn:riess} that it
is close to $-$1, showing that $\varpi$ is rather large, as we
have been assuming. We now have two relations between $\varpi$ and
L, (\ref{eq:AHH}) and (\ref{eq:ALL}), so these parameters are
fixed in terms of observed quantities. [Readers who prefer the
more traditional deceleration parameter q$_0$ will find that it is
given by
\begin{equation}\label{eq:ALLL}
\textup{q}_0 \;= \; -\;1\;+\;{{1}\over
{\varpi}}\;\textup{sec}^2({{\textup{T}}\over {\varpi
\textup{L}}}),
\end{equation}
so that q$_0$ = ${{1}\over{\;2}}(1\;+\;3$w$_0$), which is
essentially just the Raychaudhuri equation for a FRW cosmology
with flat spatial sections.]

The metric $g^-(\varpi,A)$ gives an accurate picture of spacetime
during the period when the expansion has diluted ordinary matter
and radiation to insignificance; that is, during the
``quintessence-dominated era" and during the subsequent era when
even quintessence is diluted and the geometry is dominated by the
\emph{negative} cosmological constant in the background. As we are
ignoring ordinary matter and radiation here, $g^-(\varpi,A)$
cannot of course be expected to describe the present state of the
Universe, and so at this stage it would be pointless to try to
compute L and $\varpi$ from current observational data.
\begin{figure}[!h]
\centering
\includegraphics[width=0.7\textwidth]{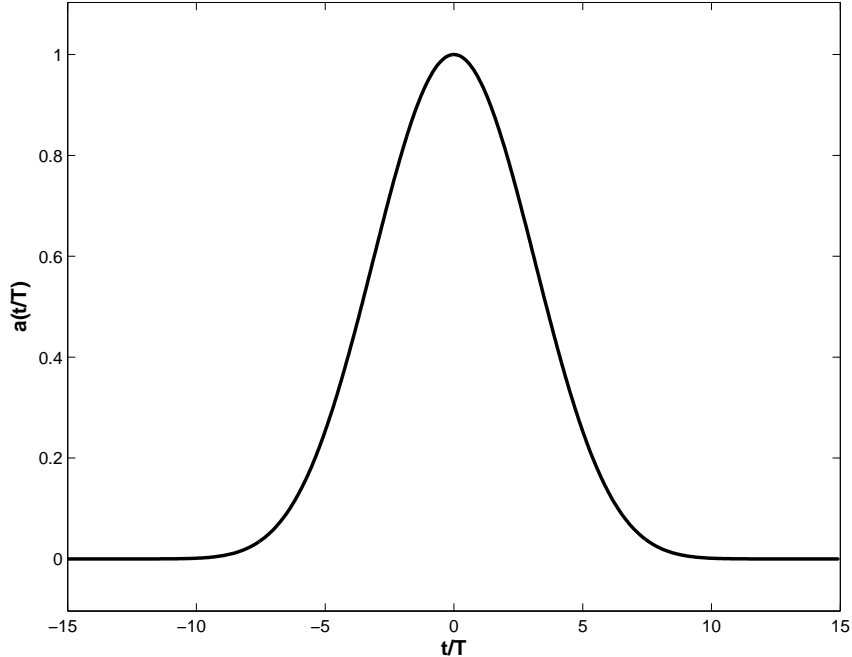}
\caption{Quintessential MM scale function for L = T, $\pi\varpi$ =
30}
\end{figure}
Purely for illustrative purposes we shall take it that L = T and
that $\pi\varpi$ = 30; that is, the total lifetime of the Universe
is assumed to be 30 times its current age. Then the graph of the
scale function against t/T is shown in Figure 2. With such values
of $\varpi$, we see that the history of this universe is as
follows.

Starting from the Big Bang at t = $-\pi\varpi$L/2, the SEC is
immediately violated and the universe accelerates. [In reality of
course there should be a period of deceleration --- now observed
\cite{kn:riess} --- but the absence of this early deceleration
from this cosmology is simply due to our neglect of matter and
radiation.] The expansion proceeds beyond the present time [t/T =
$-$14 in Figure 2] at a sedate pace, despite the fact that w is
only slightly larger than $-$1 [equation (\ref{eq:AL})], until a
large fraction of its total lifetime has passed. [This happens
because, during this time, the scale factor is still very small,
so the quintessence has not had the opportunity to cause a rapid
expansion: in Figure 2, the value of the scale function at the
present time is roughly 4.31 $\times$ 10$^{-10}$.] This
corresponds to the period in which, according to \cite{kn:riess},
we now find ourselves, with w close to $-$1 [it is equal to about
$-$0.9294 at the present time under the conditions assumed in
Figure 2] and changing only very slowly. [See \cite{kn:sahni} for
an alternative view of the observational data.] At some point,
however, the rate of expansion increases dramatically; but w
begins to rise until it reaches $-$1/3. [At this time, the scale
factor is given, according to equations (\ref{eq:AG}) and
(\ref{eq:AL}), by $(1\;-\;{{1}\over{\varpi}})^{\varpi/2}$, which
is about 1/$\sqrt{e}$ for large values of $\varpi$.] This is the
point of transition from acceleration to deceleration. The
equation-of-state parameter soon becomes positive and in fact
larger than $+$1; in this respect the situation is analogous to
the one considered in \cite{kn:turok}, though of course that work
is concerned with contraction to a minimum size instead of
expansion to a maximum. The expansion halts at t = 0, and then a
rapid contraction, still under the influence of the negative
cosmological constant in the background, begins. There is then
another transition back to an SEC-violating regime; although the
period during which the universe is very large and decelerating is
very short, by this time the universe is contracting so rapidly
that a Big Crunch cannot be averted; it takes place at t =
$\pi\varpi$L/2. It is interesting that the Universe ends its days
in a futile effort, by accelerating again, to avert destruction.
Observers at that time might be misled into believing that they
live in a de Sitter-like world which might ``bounce" and
re-expand. The key point here is that only a brief period of
deceleration is necessary to bring about a Crunch.

The causal structure of these spacetimes is rather interesting and
relevant to our later discussion, so we consider it briefly. If we
unwrap the spatial sections, so that they are copies of $\bbr^3$
instead of tori, then our spacetimes are conformal to Minkowski
spacetime. Defining a parameter $\lambda$ by cos(t/$\varpi$L) =
sech($\lambda$), we see that the full extent of conformal time is
given by
\begin{equation}\label{eq:GGGGGGG}
\varpi
\textup{L}\int_{-\infty}^{\infty}\textup{cosh}^{\varpi\;-\;1}(\lambda)d\lambda,
\end{equation}
which converges if $\varpi \;<\;1$. The spatial sections being
compact, the Penrose diagram in this relatively uninteresting case
will be rather like the one given in Figure 1 above. Since $\varpi
\;<\;1$ implies that the Strong Energy Condition is satisfied at
all times, the similarity to that case is perhaps not very
surprising. [The precise shape of the diagram in this case is,
however, at our disposal, since the scale of the spatial tori can
be fixed independently of all other parameters. Thus, the Penrose
diagram can be either short or tall, just as we decide.]

In the much more interesting case $\varpi \;>\;1$, in which the
SEC does not hold at all times, the situation is quite different.
The peculiar feature here is that both the Bang and the Crunch are
infinitely remote in conformal time. This means that the Penrose
diagram resembles that of Minkowski spacetime, but a singularity
occupies all of the region t = $\pm\infty$ on the edges of the
usual diamond, so that there are separate singularities for
timelike and null geodesics. Thus the Bang is visible at all
times, but the part of the Bang that one can see is not the part
from which timelike geodesics emanate.

If we now compactify the spatial sections to tori, then there is
no longer any spatial infinity or any null infinity. There is
still an infinity for timelike geodesics, of course, but it is
represented only by a pair of points in the diagram, one each for
the Bang and the Crunch, shown as the heavy dots in Figure 3.
Essentially what happens is that null geodesics wind around the
torus an infinite number of times as the Universe reaches the
Crunch [or as they are traced back to the Bang]; they cannot reach
either singularity. In fact, the spacetime is null geodesically
complete. Thus the Bang itself is dark.

 Since a torus is not globally isotropic, it is difficult
to represent the situation in a fully adequate way on a Penrose
diagram, but we can obtain a partly satisfactory picture by
focusing on one of the coordinate directions
$\theta_1$,$\theta_2$, or $\theta_3$; let us pick $\theta_1$. The
vertical straight line in Figure 3 represents $\theta_1$ = 0, and
the two curved lines correspond to $\theta_1$ = $\pm\pi$.
\begin{figure}[!h]
\centering
\includegraphics[width=0.15\textwidth]{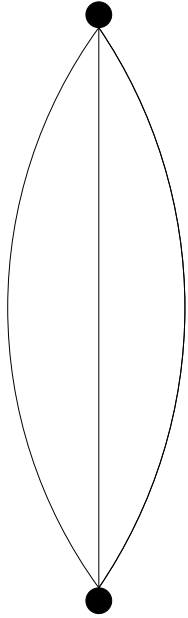}
\caption{Penrose diagram of the Quintessential MM Cosmologies}
\end{figure}
[That is, they indicate the topological identifications which
define a torus. One can also, however, think of them as the
timelike geodesic traced out by a point on a suppressed two-torus,
which, like any other timelike geodesic, begins in the Bang and
ends in the Crunch.]

One strange property of this cosmology is that null geodesics can
circumnavigate the Universe, no matter how large the spatial tori
may be. Indeed, at any point in this spacetime one can find a null
geodesic which has performed an arbitrarily large number of
circumnavigations. This is in sharp contrast to the case of Figure
1; in that spacetime, circumnavigations are completely impossible
[since the diagram is less than twice as high as it is wide]. In
principle, therefore, this cosmology predicts that the non-trivial
topology of its spatial sections must be visible, and this appears
to be in conflict with the observations \cite{kn:cornish}.
However, the cosmic microwave radiation which we see today does
not emanate from the Bang itself: one can think of the
corresponding null geodesics as beginning on an initial spacelike
hypersurface. If the present age of the Universe is sufficiently
short relative to its total lifetime, and if the parameter A in
(\ref{eq:AH}) is chosen to be sufficiently large, then one can
easily arrange for the past lightcone of the present moment to
intersect this initial surface before the cone extends far enough
to reveal the topology. Notice in this connection that there is no
``isotropy problem" here, since the Bang is a single point in the
Penrose diagram.

To summarize: we have a family of cosmological models which are
obtained in much the same way as those considered by Maldacena and
Maoz, but with flat spatial sections and with a transition from
acceleration to deceleration, culminating in a Big Crunch. They
seem to be physically acceptable, within the constraints imposed
by the fact that we have not tried to set up a fully realistic
matter model. Nevertheless the global structure of the Lorentzian
version differs very radically from that of the Maldacena-Maoz
spacetime [Figure 1]. We shall return to this in the Conclusion.

However, the most puzzling feature of these spacetimes only
appears when we turn to the Euclidean version of (\ref{eq:AH}),
given by
\begin{equation}\label{eq:AM}
g^+(\varpi,A) = dt^2 + A^2\;\textup{cosh}^{2\varpi}({{t}\over
{\varpi \textup{L}}})\;[d\theta_1^2 + d\theta_2^2 + d\theta_3^2].
\end{equation}
For large positive and negative t this is, for all $\varpi$,
approximately
\begin{equation}\label{eq:SPLAT}
g^+(\varpi,A)\; \approx \; dt^2 + \;
4^{-\varpi}A^2\;\textup{exp}(2|t|/\textup{L})\;[d\theta_1^2 +
d\theta_2^2 + d\theta_3^2],
\end{equation}
precisely as Maldacena and Maoz require [see equation
(\ref{eq:D})]. Clearly $g^+(\varpi,A)$ is defined on a space with
a conformal boundary which is compact but which consists, at least
in the simplest interpretation, of two disconnected tori, just as
the conformal boundary of the Euclidean Maldacena-Maoz space
consists of two disconnected spheres. The great difference between
the two only becomes apparent when we compute the eigenvalues of
the Ricci tensor of $g^+(\varpi,A)$. They are given by
\begin{equation}\label{eq:AN}
\textup{Ric}_{(0)}(g^+(\varpi,A))\;=\;-\;{{3}\over{\textup{L}^2}}\;+\;{{3}\over{\textup{L}^2}}\;[1\;-\;{{1}\over
{\varpi}}]\;\textup{sech}^2({{t}\over {\varpi \textup{L}}})
\end{equation}
and
\begin{equation}\label{eq:AO}
\textup{Ric}_{(1,2,3)}(g^+(\varpi,A))\;=\;
-\;{{3}\over{\textup{L}^2}}\;+\;{{3}\over{\textup{L}^2}}\;[1\;-\;{{1}\over
{3\varpi}}]\;\textup{sech}^2({{t}\over {\varpi \textup{L}}}).
\end{equation}
At once we see that our quintessence field [which always satisfies
the inequality (\ref{eq:HHH})] can never make a negative
contribution to the last three eigenvalues. It \emph{can} make a
negative contribution to Ric$_{(0)}(g^+(\varpi,A))$, \emph{but
only in the cosmologically least interesting case}, namely, when
$\varpi \; < \;$ 1. [The very fact that this is possible confirms
that we are indeed dealing with a generalized axion here: see the
discussion in the previous section.]

In the case of real interest, $\varpi\; > \;$ 1, the Witten-Yau
inequalities (\ref{eq:F}) are fully satisfied; \emph{and yet the
Euclidean boundary is disconnected.} This seems to contradict the
Giddings-Strominger requirement \cite{kn:GS} that a wormhole
should make a negative contribution to the Euclidean Ricci
curvature, but this is readily understood when we observe that
these cosmological wormholes are not asymptotically flat. Less
easy to understand is the apparent contradiction of the Witten-Yau
theorem. We now explain how this is possible.

\addtocounter{section}{1}
\section*{6. Escaping the Menace of the WY Theorem, Part 2}

In order to understand this puzzling situation, let us remind
ourselves of the definition of a conformal boundary.

Let M$^{n+1}$ be a non-compact (n+1)-dimensional manifold which
can be regarded as the interior of a compact, connected
manifold-with-boundary ${\overline \textup{M}}^{n+1}$, and let
N$^n$ be the boundary (which need not be connected).  Let $g^+$(M)
be a smooth Euclidean metric on M$^{n+1}$ such that there exists a
function G on ${\overline \textup{M}}^{n+1}$ with the following
properties:

\medskip

[a] G(x) = 0 if and only if x $\in$ N$^n$;

\medskip

[b] dG(x) $\neq$ 0 for all x $\in$ N$^n$;

\medskip

[c] G$^2$$g^+$(M) extends continuously to a metric on
            ${\overline \textup{M}}^{n+1}$;

\medskip

[d] If $|\;$dG$\;|$ is the norm of dG with respect to the extended
metric, then $|\;$dG$\;|$, evaluated on N$^n$, must not depend on
position there.

\medskip

Then we say that ${\overline \textup{M}}^{n+1}$ is a conformal
compactification of M$^{n+1}$, that N is the conformal boundary,
and that G is a defining function for N. [It is customary to
impose various differentiability conditions on G, but we shall
only require the existence of continuous first derivatives.]

 The first three conditions ensure that the boundary is
infinitely far from any point in the interior as measured by
$g^+$(M). The last point is not usually mentioned, because it is
not necessary when the bulk is an Einstein space; however, we do
need it here. It ensures that all sectional curvatures along
geodesics ``tending to infinity'' approach a common negative
constant, the asymptotic sectional curvature, which is equal to
$-\;|\;$dG$\;|^2$, evaluated on N$^n$. [Despite appearances, this
is independent of the choice of G.] The metric is said to be
\emph{asymptotically hyperbolic} for this reason. It follows that
the eigenvalues of the Ricci tensor must all approach
$-\;3|\;$dG$\;|^2$ in four dimensions.

For $g^+(\varpi,A)$, a natural choice for G can be constructed as
follows. First define a constant c$_{\varpi}$ by
\begin{equation}\label{eq:XXXX}
\textup{c}_{\varpi}\; =
\;{{\varpi}\over{\pi}}\int_{0}^{\infty}\textup{sech}^{\varpi}(\zeta)d\zeta
\;;
\end{equation}
the integral clearly converges, so c$_{\varpi}$ is well defined;
note that it depends only on $\varpi$. Now define a new coordinate
$\theta$ by c$_{\varpi}$Ld$\theta$ =
$\pm$sech$^{\varpi}$(${{t}\over {\varpi \textup{L}}}$)dt, where
the sign is chosen as + when t is positive, $-$ when t is
negative. An elementary calculation shows that the range of
$\theta$ is just $-\pi$ to $+\pi$, corresponding to t ranging from
$-\infty$ to $+\infty$; that is, the boundaries are at the finite
$\theta$ values $\pm\pi$, and $\theta$ = 0 corresponds to t = 0.
Now solve for t in terms of $\theta$ and use this to express
sech$^{\varpi}$(${{t}\over {\varpi \textup{L}}}$) in terms of
$\theta$. The resulting function, denoted G$_{\varpi}(\theta)$,
vanishes at $\pm\pi$, is at least once differentiable, and
satisfies all of the conditions for a defining function for
$g^+(\varpi,A)$. For example, the reader can verify that
G$_2$($\theta$) is given simply by 1$\; - \;(\theta/\pi)^2$ and
that it is a defining function for $g^+(2,A)$.

Using $\theta$, we find that in the general case our metric is
\begin{equation}\label{eq:AMM}
g^+(\varpi,A) =
\textup{G}^{-2}_{\varpi}(\theta)\;[c_{\varpi}^2\textup{L}^2d\theta^2
\; +\; A^2(d\theta_1^2 \;+\; d\theta_2^2 \;+\; d\theta_3^2)].
\end{equation}
Computing the norm of dG$_{\varpi}$ with respect to
G$_{\varpi}^2(\theta)$$g^+(\varpi,A)$ [using of course the inverse
metric to evaluate the norm of a one-form], we obtain
\begin{equation}\label{eq:AP}
-\;|\;\textup{dG}_{\varpi}|^2\;=\;-\;{{1}\over{\textup{L}^2}}[1\;-\;\textup{G}_{\varpi}^{2/\varpi}(\theta)].
\end{equation}
All of the above conditions are satisfied and the asymptotic
sectional curvature [towards both connected components of the
conformal boundary] is $-$1/L\2; the Ricci eigenvalues must
approach $-$3/L\2, which indeed they do.

Now the Witten-Yau inequalities (\ref{eq:F}) have a
straightforward meaning when the eigenvalues are \emph{constants},
as of course they are when the bulk is an Einstein manifold. But
when, as in the case considered by Maldacena and Maoz, the bulk is
not an Einstein manifold, the Ricci eigenvalues are functions of
position. It turns out that, in this situation, the Witten-Yau
theorem needs not just (\ref{eq:F}) but also some restriction on
the rate at which the Ricci eigenvalues approach their asymptotic
value, $-$3/L\2. This restriction was supplied by Cai and Galloway
\cite{kn:galloway}, and may be stated simply as follows in the
general case.

Cai and Galloway express the metric of an asymptotically
hyperbolic space, near to any connected component of the boundary,
in the form
\begin{equation}\label{eq:AV}
g^+(\textup{M})\; = \;{{\textup{L}^2}\over{r^2}}[dr^2\;+\;g_r].
\end{equation}
The advantage of this way of writing the metric is that the
coordinate r measures distance to the boundary according to the
re-scaled metric [so that the boundary is at r = 0]. Using r, we
can measure the rate at which a given function J(r) tends to zero
as the boundary is approached
--- clearly J(r) may be said to tend to zero very
rapidly towards infinity if J(r)/r$^n$ tends to zero towards the
boundary even for some large power n. In particular, suppose that
the asymptotic sectional curvature is $-$1/L\2, so that the Ricci
eigenvalues satisfy Ric$_{(j)}\;+\;3/\textup{L}^2\;\rightarrow\;$0
as conformal infinity is approached. The question now is \emph{how
quickly} these functions of position tend to zero.

We shall say that the \emph{Cai-Galloway conditions}
 are satisfied for the conformal
compactification of a four-manifold if
\begin{equation}\label{eq:AQ}
\textup{r}^{-2}[\textup{Ric}_{(j)}\;+\;3/\textup{L}^2]\;\rightarrow\;0
\end{equation}
uniformly as infinity is approached, for all j. Cai and Galloway
show, in a beautiful paper \cite{kn:galloway} using quite
different techniques to those of \cite{kn:yau}, that if all of the
other conditions of the Witten-Yau theorem are satisfied,
including the inequalities (\ref{eq:F}), \emph{and if the
Cai-Galloway conditions are also valid}, then the conformal
boundary must be connected.

Now in the case of the quintessential manifolds we have been
considering, the metric $g^+(\varpi,A)$ can always be put in the
form (\ref{eq:AV}) near either component of the boundary by
defining r by t = $\pm$L$\;ln$(r/L) [the choice of sign being
determined by which component of the boundary we select]; for then
we can express the defining function G$_{\varpi}$($\theta$) in
terms of r as
\begin{equation}\label{eq:AW}
\textup{G}_{\varpi}\;=\;2^{\varpi}\;{{r}\over{\textup{L}}}\;[1\;+\;({{r}\over{\textup{L}}})^{2/\varpi}]^{-\varpi}.
\end{equation}
Now equations (\ref{eq:AN}) and (\ref{eq:AO}), expressed in terms
of r, yield
\begin{equation}\label{eq:AR}
r^{-2}[\textup{Ric}_{(0)}(g^+(\varpi,A))\;+\;{{3}\over{\textup{L}^2}}]\;=\;{{12\;[\varpi\;-\;1]\;(r/\textup{L})^{{{2[1\;-\;\varpi]}\over{\varpi}}}}\over{\varpi\textup{L}^4[1\;+\;(r/\textup{L})^{{{2}\over{\varpi}}}]^2}},
\end{equation}
\begin{equation}\label{eq:AS}
r^{-2}[\textup{Ric}_{(1,2,3)}(g^+(\varpi,A))\;+\;{{3}\over{\textup{L}^2}}]\;=\;{{12\;[\varpi\;-\;{{1}\over{3}}]\;(r/\textup{L})^{{{2[1\;-\;\varpi]}\over{\varpi}}}}\over{\varpi\textup{L}^4[1\;+\;(r/\textup{L})^{{{2}\over{\varpi}}}]^2}}.
\end{equation}
At once we see where the problem lies. Recalling that r tends to
zero towards the boundary, we see that for values of $\varpi$
strictly between 1/3 and 1, the Cai-Galloway conditions are
satisfied [the right hand sides tend to zero] but the Witten-Yau
inequalities are not. For values of $\varpi$ greater than or equal
to unity, the reverse is true. In short, there is no choice of
$\varpi$ which can satisfy all of the conditions required by
Witten, Yau, Cai, and Galloway. \emph{This is how the Euclidean
boundary can be disconnected for all of these cosmologies}: the
contribution made by [``axionic"] quintessence to the Euclidean
Ricci curvature is either of the wrong sign or it decays towards
the boundary too slowly for the Witten-Yau-Cai-Galloway theorem to
apply.

For the Maldacena-Maoz metric (\ref{eq:GG}) we can again choose a
new coordinate such that the boundary components are at finite
values of the new coordinate, and then express
$[(\alpha+{{1}\over{4}})^{1/2}\textup{cosh}(2t/\textup{L})-{{1}\over{2}}]^{-1/2}$
in terms of this new coordinate to obtain a defining function.
Again $g_{MM}$ can be written in the form (\ref{eq:AV}) near
either component of the boundary, and one can show, using equation
(\ref{eq:PP}) and the tracelessness of the energy-momentum tensor,
that
\begin{equation}\label{eq:AT}
r^{-2}[\textup{Ric}_{(0)}(g_{MM}) \;+\; {{3}\over
{\textup{L}^2}}]\; = \;{{\;-\; 3\alpha
r^2}\over{\textup{L}^6}\;[{{1}\over{2}}(\alpha+{{1}\over{4}})^{1/2}\;\{1\;+\;({{r}\over{\textup{L}}})^4\}\;-\;{{r^2}\over{2\textup{L}^2}}]^{2}}
\end{equation}
\begin{equation}\label{eq:AU}
r^{-2}[\textup{Ric}_{(1,2,3)}(g_{MM}) \;+\; {{3}\over
{\textup{L}^2}}]\; = \;{{\alpha
r^2}\over{\textup{L}^6}\;[{{1}\over{2}}(\alpha+{{1}\over{4}})^{1/2}\;\{1\;+\;({{r}\over{\textup{L}}})^4\}\;-\;{{r^2}\over{2\textup{L}^2}}]^{2}}.
\end{equation}
That is, the effect of the matter is of order r$^4$ towards
infinity, a result which is not surprising in view of the fact
that we are dealing with Yang-Mills matter. The conditions
(\ref{eq:AQ}) are always satisfied in this case; in fact the rate
of decay is more than sufficiently rapid to satisfy the
Cai-Galloway conditions. Thus the Maldacena-Maoz space always
[that is, for all values of $\alpha$] violates the Witten-Yau
inequalities and satisfies the Cai-Galloway conditions. [In fact,
the behaviour of the Maldacena-Maoz space is very similar to that
of the quintessential space with $\varpi$ = 1/2, which also has a
factor proportional to r$^2$ on the right hand sides of
(\ref{eq:AR}) and (\ref{eq:AS}).]

To understand these results physically, notice first that the
Maldacena-Maoz space and the quintessential space with $\varpi <
1$ violate the Witten-Yau inequalities precisely because they
satisfy the SEC. Conversely, the quintessential spacetimes do
temporarily violate the SEC [while still having a Bang and a
Crunch] if the parameter $\varpi$ exceeds unity, and this is
exactly the condition which ensures that the Witten-Yau
inequalities are satisfied in the Euclidean case. In short, the
observation of cosmic acceleration means that the Witten-Yau
inequalities actually \emph{are} a physically reasonable set of
conditions in cosmology, not because they correspond to the SEC
\emph{but because they correspond to its violation}.

On the other hand, the Cai-Galloway conditions are also physically
well-justified \emph{in some circumstances}. The theory of
asymptotically anti-de Sitter spacetimes was developed in
\cite{kn:magnon}, in a way which generalises the usual theory of
asymptotically flat spacetimes. The fall-off conditions for matter
given there are that $\Omega^{-3}$T$^i_j$, where $\Omega$ is a
canonical conformal factor which tends to zero towards infinity
and T$^i_j$ is the (1,1) energy-momentum tensor, should have a
smooth [not necessarily zero] limit at the boundary. This
corresponds to requiring that the Ricci eigenvalues [in the
Euclidean version] should tend to zero at least as quickly as
r$^2$ [since we are requiring the limit to be zero]. Thus we see
that the Cai-Galloway conditions are in fact a Euclidean version
of those normally employed in asymptotically anti-de Sitter
physics. Thus the Cai-Galloway conditions are well-motivated in
such situations.

However, \cite{kn:magnon} is explicitly concerned with
generalizations of asymptotically flat spacetimes --- that is, the
applications intended are to black holes and similar localised
phenomena. In cosmology we expect rates of decay towards
boundaries which are slower, not faster, than the analogous rates
for black holes. Indeed, quintessence is the canonical example of
a form of matter which decays more slowly as the Universe expands
than any ordinary form of matter. Recall that, in a FRW cosmology
with a constant equation-of-state parameter w and a scale factor
a, the density decays according to a$^{-\;3(1\;+\;w)}$. Cosmic
acceleration requires w $< -\;{{1}\over{3}}$, and so we see that,
at least in the constant-w case, the condition that the Universe
should accelerate is that the density should decay at a
\emph{maximum} rate of a$^{-2}$. Transferring this to the
Euclidean domain, and comparing the FRW metric with (\ref{eq:AV}),
we see that \emph{we should not expect the Cai-Galloway conditions
to be satisfied by the Euclidean version of an accelerating
cosmology}, though they should be perfectly reasonable for [say]
Euclidean versions of AdS black holes.

Thus we have finally uncovered the real underlying reason for the
ability of our anti-de Sitter-like quintessence cosmologies to
have Euclidean versions with disconnected boundaries: it is that
the case where these spacetimes  accelerate is precisely the case
where the Euclidean versions violate the Cai-Galloway conditions.
The quintessence simply decays towards the boundaries ``too
slowly".\footnote{This slow decay is also reflected in the fact
that, for large values of $\varpi$, the second and higher
derivatives of the function G$_{\varpi}$ do not extend to the
boundary. It is possible that the Cai-Galloway theorem does not
apply to such a case; if not, this is another way of relating the
slow decay to the disconnectedness of the boundary. The author is
grateful to Professor Galloway for very helpful correspondence on
this and for correcting the statement of the Cai-Galloway theorem
in an earlier draft.}

We can summarize as follows. Cai and Galloway \cite{kn:galloway}
showed that, in the non-Einstein case, connected Euclidean
boundaries can be ensured if the Witten-Yau inequalities are
satisfied, \emph{provided} that the functions
Ric$_{(j)}\;+\;3/\textup{L}^2$ decay towards infinity sufficiently
rapidly. But in our anti-de Sitter quintessence cosmologies these
functions do not decay so rapidly in the physically most
interesting cases. Thus disconnected Euclidean boundaries cannot
be excluded, and in fact they do occur. More generally, we can
expect that

\medskip

[a] The Witten-Yau inequalities \emph{will} be satisfied by the
Euclidean version of an anti-de Sitter-like cosmology which
undergoes periods of acceleration, as our Universe does.

\medskip

[b] Such cosmologies will generically have Euclidean versions
which violate the Cai-Galloway conditions; and this violation will
permit [though not require] a disconnected conformal boundary.

\medskip

In other words, it seems that \emph{the obstruction to
disconnected
 boundaries discovered by Witten and Yau is naturally evaded in the version
  of cosmological holography proposed by Maldacena and Maoz}.

 \addtocounter{section}{1}
\section*{7. Conclusion}

Our objective in this work is to persuade the reader that
situations like the one considered by Maldacena and Maoz
\cite{kn:maldacena}, where the holographic picture [apparently]
involves two boundaries but only one bulk, can in fact arise in a
quasi-realistic cosmological setting obtained by introducing dark
energy into AdS\4. We have tried to do this by identifying
precisely how the quintessential spacetime [with metric given in
equation (\ref{eq:AH})] is able to evade the Witten-Yau theorem
and its extension due to Cai and Galloway \cite{kn:galloway}. The
essential point turns out to be the fact that in cosmology one is
not always entitled to prescribe very rapid asymptotic rates of
decay of matter fields towards infinity. Dark energy, in
particular, dilutes with the cosmic expansion very slowly, too
slowly for the theorem to apply. In essence, it is the peculiar
refusal of dark energy to dilute in the conventional way that
permits the Euclidean version of spacetime to have a disconnected
boundary.

The present work is intended to be complementary to
\cite{kn:maldacena}. The latter was concerned with showing that it
is possible to have a well-behaved field theory, on a disconnected
boundary, which is dual to a known truncation of a compactified
supergravity theory in the bulk. This is an important point, but
it leads to cosmological models with no acceleration. Here we have
a bulk, again with a disconnected Euclidean boundary, which leads
to a more realistic cosmology, but we lack as yet a dual
description of the bulk matter. The fact that the Euclidean
version of our quintessence field reveals it to be a ``generalized
axion" is probably relevant here. Perhaps the ideas advanced in
\cite{kn:townsend}\cite{kn:bak}\cite{kn:jarv} may prove useful. It
would also be valuable to have more examples of \emph{exact}
quintessence spacetime metrics [see the methods of
\cite{kn:russo}, see also \cite{kn:ish2}] leading to Euclidean
spaces with multiple boundaries. It may be possible to learn
something useful from manifolds with \emph{more} than two boundary
components.

Granting, as now seems likely, that there really are no physical
or mathematical objections to Maldacena-Maoz cosmologies, one has
to confront the original issue identified by Witten and Yau
\cite{kn:yau}: how can two apparently independent field theories
[on the boundary] both be dual to the same bulk? It seems that
either the two field theories, despite appearances, are not really
independent [see for example \cite{kn:minic}], or the bulk,
despite appearances, is not really connected. The latter is hard
to believe here, however, simply because the only natural place
for the spacetime to split is at its point of maximum expansion,
precisely when it is ``most classical". Other, less obvious
solutions of this puzzling problem are discussed in the conclusion
of \cite{kn:maldacena}.

We close with some speculative remarks related to this issue.

First, the reader should be aware that the statement that
equations (\ref{eq:GG}) and (\ref{eq:AM}) represent metrics
defined on a space with \emph{two} boundaries is not a
mathematical fact; it is an interpretation of the structure of the
metric. To understand this, recall that there is a natural choice
of coordinate $\theta$, running from $-\pi$ to $+\pi$, such that
(\ref{eq:AM}) can be re-expressed in the form (\ref{eq:AMM}). For
simplicity let us choose the radii of the three-tori to be given
by A = c$_{\varpi}$L. Then (\ref{eq:AMM}) becomes
\begin{equation}\label{eq:YECCH}
g^+(\varpi,A)\; =\;
c^2_{\varpi}\textup{L}^2\textup{G}_{\varpi}(\theta)^{-2}\;[d\theta^2
\;+\; d\theta_1^2 \;+\; d\theta_2^2 \;+\; d\theta_3^2].
\end{equation}
The ``obvious" interpretation of \emph{this} is that the Euclidean
version of the quintessential metric is defined on an open subset,
corresponding to the open interval ($-\pi$, $+\pi$) for $\theta$,
in the four-dimensional torus S$^1\;\times\;$T\3 = T$^4$. A
conformal transformation which strips away the factor
$c^2_{\varpi}\textup{L}^2\textup{G}_{\varpi}(\theta)^{-2}$ allows
us to extend the metric to the usual one on the cubic four-torus
with all four circumferences equal to 2$\pi$. [Other choices of A
simply yield non-cubic tori.] The conformal factor tends to
infinity as we approach either $\theta$ = $-\pi$ or $\theta$ =
$+\pi$, \emph{but this is one three-torus, not two}; it is the
same three-torus being approached from opposite sides. Similarly,
the Euclidean Maldacena-Maoz metric can be regarded as being
defined on an open submanifold of S$^1\;\times\;$S\3 or
S$^1\;\times\;\bbr$P\3: we have from (\ref{eq:GG})
\begin{equation}\label{eq:GGA}
g^+_{MM}\; = \;
\textup{G}_{MM}(\psi)^{-2}\;[\textup{b}_{\alpha}^2\textup{L}^2\;d\psi^2
+ \textup{L}^2(d\chi^2 + \textup{sin}^2(\chi)\{d\theta^2 +
\textup{sin}^2(\theta)d\phi^2\})],
\end{equation}
where b$_{\alpha}$ is a constant, depending only on $\alpha$,
defined as
\begin{equation}\label{eq:XXXXX}
\textup{b}_{\alpha}\; =
\;{{1}\over{2\pi}}\int_{0}^{\infty}{{d\zeta}\over{\sqrt{(\alpha+{{1}\over{4}})^{1/2}\textup{cosh}(\zeta)-{{1}\over{2}}}}}\;,
\end{equation}
where $\psi$ is an angular coordinate on the circle S$^1$, ranging
between $\pm\pi$, defined by \\ b$_{\alpha}$Ld$\psi$ =
$\pm[(\alpha+{{1}\over{4}})^{1/2}\textup{cosh}(2t/\textup{L})-{{1}\over{2}}]^{-1/2}$dt,
and where G$_{MM}$($\psi$) is the defining function. [Notice that
b$_{\alpha}$ and [therefore] $\psi$ are well-defined \emph{unless}
$\alpha$ is exactly zero.] Again the conformally deformed version
extends to \emph{all} of S$^1\;\times\;$S\3 or
S$^1\;\times\;\bbr$P\3, with the circle being of radius
b$_{\alpha}$L.

Of course this means that we are thinking of the conformal
compactification space as a compact \emph{manifold} instead of a
compact manifold-with-boundary; here, infinity corresponds to a
special submanifold instead of a pair of boundary components. [It
is a simple exercise to adapt the usual definition of a conformal
compactification to a definition in terms of infinitely distant
\emph{submanifolds} instead of \emph{boundaries}.]

The point to be stressed is that one cannot \emph{prove}
mathematically that either of these interpretations --- ``two
boundaries" or ``one submanifold" --- is the ``correct" one.
Equation (\ref{eq:AMM}) is after all obtained from (\ref{eq:AM})
simply by changing a coordinate, and (\ref{eq:YECCH}) is just a
special case of (\ref{eq:AMM}). The question as to whether there
really are two field theories here is a physical, not a
mathematical one: it cannot be decided simply by inspecting the
metric. If (\ref{eq:AMM}) had been found first, then we might have
argued that the ``toral" interpretation which it suggests is more
natural than the ``cylindrical" one suggested by (\ref{eq:AM}),
and then the ``double-boundary" conundrum \emph{would never have
arisen}. Note that the boundary of \emph{every} compact
manifold-with-boundary can be interpreted as a submanifold of a
compact \emph{manifold}: simply take two copies of the
manifold-with-boundary and identify them along the boundary. Thus
the ``submanifold" interpretation of conformal infinity is quite
as general as the more familiar ``boundary" interpretation.
Indeed, in the case of the Euclidean Maldacena-Maoz space with S\3
sections, we can see how this works explicitly: writing the metric
as in equation (\ref{eq:GGA}), let $\alpha$ tend to zero. Then
from equation (\ref{eq:XXXXX}), this causes b$_{\alpha}$ to
diverge: the circle ``snaps" as the wormhole pinches off and we
are left with two copies of Euclidean AdS\4 identified along a
common boundary. [However, this would lead to a new problem: two
bulks for one field theory.]

These observations allow us to formulate the ``double-boundary
problem" in a more physical way: the problem is connected with the
question as to whether we can find a physical reason for
preferring manifolds-with-boundary to manifolds. The situation
here is very much analogous to the interpretational ambiguities of
the Randall-Sundrum models \cite{kn:randall1}\cite{kn:randall2}:
does the bulk extend away from the brane-world on both sides, or
only on one? In the first case, the brane is a submanifold and
\emph{not} a boundary, as it is in the second case. If further
investigations of brane-world models suggest that branes at ``the
end of the world" are physically unacceptable, then this might
well extend to a physical argument showing that the ``circular"
interpretations of $g^+(\varpi,A)$ and $g^+_{MM}$ suggested by
(\ref{eq:AMM}) and (\ref{eq:GGA}) are the correct ones. This
prompts a difficult question: what would this conclusion imply for
the Lorentzian versions of these spaces?

As is well known, the conformal compactification of Lorentzian de
Sitter spacetime also has an apparently double boundary, and this
is the case also in asymptotically de Sitter ``bouncing"
cosmological models. The four-dimensional de Sitter metric can be
written in globally valid coordinates as
\begin{eqnarray}\label{eq:dS}
g(dS_4) = {{L^2} \over {\textup{cos}^2(\psi/2)}}\;[-
{{1}\over{4}}\;d\psi^2 + d\chi^2 + \textup{sin}^2(\chi)\{d\theta^2
+ \textup{sin}^2(\theta)d\phi^2\}],
\end{eqnarray}
where $\psi$ is again an angular coordinate running from $-\pi$ to
$+\pi$. The fact that the conformal factor is a periodic function
makes a topological identification of the boundary components,
again converting a manifold-with-boundary to a compact manifold,
particularly natural from a mathematical point of view.
Physically, too, the identification seems natural, because the
rapid contraction/expansion of the universe in such models at very
early/late times does indeed tend to render physical conditions
identical towards both components of infinity. [Such an
identification of the conformally related spacetime does not, of
course, violate causality in the original spacetime, since the
closed timelike worldlines are infinitely long there.]

In the cases discussed here, however, it is hard to use such
arguments because the boundaries we are discussing are boundaries
of the \emph{Euclidean} versions of our spacetimes. Particularly
in the case of $g^{\pm}(\varpi,A)$ with $\varpi \;
>\;$ 1, it is far from clear that it is correct to relate physical
conditions near the singularities in Figure 3 to conditions near
the boundaries of the Euclidean version. If it were correct to do
so, then identifying the connected components of the Euclidean
boundary would [presumably] entail identifying the singularities
at the top and bottom of Figure 3. While this does not violate
causality, it might do so if the singularities were somehow
resolved. Perhaps Nature relies on cosmological singularities to
preserve causality.

This discussion leads to our second observation. One striking
feature of all of the spacetimes discussed here is that they seem
to be holographic \emph{only} in their Euclidean versions. This is
in contrast to AdS\5, and to some extent also to de Sitter
spacetime dS\4. Let us explain. The conformal boundary of the
[simply connected] Lorentzian version of AdS\5 has the structure
$\bbr\;\times$ S\3, and this is a suitable background for a
Lorentzian field theory. The Euclidean version has S$^4$ as its
boundary, and this again is a suitable domain for a Euclidean
field theory. For dS\4, the situation is less satisfactory: in the
Lorentzian case, the boundary consists of two copies of
\emph{Euclidean} S\3, and this of course is part of the reason for
the fact that it is difficult to make the dS/CFT correspondence
\cite{kn:strominger} work as effectively as AdS/CFT. The Euclidean
version of dS\4 is normally taken to be the four-sphere S$^4$ [but
see \cite{kn:exploring}], which has no holographic dual whatever
since it has no boundary. When we consider the spacetimes
discussed here, we find that the Maldacena-Maoz spacetime and our
quintessential spacetime for $\varpi \;<\;1$ both have Penrose
diagrams like the one in Figure 1. These do have extended
boundaries, though not of the same kind as those of the Lorentzian
versions of AdS\5 and dS\4. However, they only have such Penrose
diagrams because they never violate the SEC. In the more realistic
case where there are periods of acceleration, that is, in the
quintessential case with $\varpi \;>\;1$, the Penrose diagram is
as in Figure 3. In this case there is no hope of establishing a
holographic duality with a boundary consisting of the two singular
points in that diagram. It would be very interesting to know
whether the situation depicted in Figure 3 is in some sense
generic for anti-de Sitter-like cosmologies which, on the one
hand, have well-behaved Euclidean versions and which, on the
other, can accommodate periods of acceleration together with a
relatively short period of rapid deceleration, as in Figure 2. If
this is so, it suggests that holography in cosmology may be a
strictly Euclidean phenomenon.

\addtocounter{section}{1}
\section*{Acknowledgements}
The author is grateful for the kind hospitality of the High Energy
Section of the Abdus Salam International Centre for Theoretical
Physics, where most of this work was done. He also wishes to thank
Greg Galloway for very helpful correspondence, Ishwaree Neupane
for useful discussions, and the referee for the observation that
our quintessence field may be some kind of axion. He is also
grateful to Soon Wanmei for the diagrams and indeed for
everything.

\end{document}